\documentclass[pnas-new,twocolumn,floatfix,amsmath,amssymb,showpacs,superscriptaddress]{revtex4-1}
\usepackage{graphicx,placeins}
\usepackage{float}
\usepackage{color}
\usepackage{dcolumn}
\usepackage{bm}
\usepackage{bbold}
\usepackage{epsfig,psfrag,amsmath,amssymb, float}
\usepackage{nccmath}
\input{epsf}
\usepackage{hyperref}
\usepackage{pifont,booktabs,siunitx}
\hypersetup{colorlinks=true,urlcolor=blue}
\usepackage[noabbrev]{cleveref}

\usepackage[percent]{overpic}

\begin{document}

\title{Spontaneous altermagnetism in multi-orbital correlated electron systems}
\author{Nitin Kaushal}
\affiliation{Department of Physics and Astronomy and Quantum Matter Institute, University of British Columbia, Vancouver, British Columbia BC V6T 1Z4, Canada}
\author{Adarsh S. Patri}
\affiliation{Department of Physics and Astronomy and Quantum Matter Institute, University of British Columbia, Vancouver, British Columbia BC V6T 1Z4, Canada}
\author{Marcel Franz}
\affiliation{Department of Physics and Astronomy and Quantum Matter Institute, University of British Columbia, Vancouver, British Columbia BC V6T 1Z4, Canada}
\date{\today}

\begin{abstract}
Altermagnets have attracted considerable attention in recent years owing to their potential technological applications in spintronics and magnonics. Recently, a new class of {\it{spontaneous altermagnets}} has been theoretically predicted in a correlated two orbital model, driven by the coexistence of antiferromagnetic spin and staggered orbital ordering, thus broadening the scope of altermagnetic phenomena to systems with strong correlations. It has been noted, however, that the required spin and orbital order violates the well-established Goodenough-Kanamori (GK) rules, which underlie much of our understanding of magnetism in complex systems. Here we show that materials with three active orbitals may offer a more realistic route to this exotic state. Specifically, we consider a two-dimensional system with $t_{2g}^{2}$ electrons and identify a novel microscopic mechanism that allows the formation of a spontaneous altermagnetic Mott insulator.  We explain how the GK rules are circumvented and provide the stability criteria by employing unbiased mean-field and density matrix renormalization group calculations.
In addition, for the first time, we uncover the presence and microscopic origin of chirally split magnons in these spontaneous altermagnets, with experimentally measurable spin conductivities. Finally, we predict that the application of a small in-plane magnetic field induces, in the presence of weak atomic spin-orbit coupling, an as-yet unreported {\it hybrid chiral magnon–orbiton mode} with a non-zero orbital polarization giving rise to finite longitudinal and transverse orbital conductivities under a thermal gradient.

\end{abstract}
\maketitle

\section{Introduction}
The interplay between charge, spin, and orbital degrees of freedom in transition metal-based materials~\cite{Tokura01} has led to a variety of interesting phenomena such as unconventional superconductivity~\cite{Si01,Daghofer01}, colossal magnetoresistance~\cite{DagottoCMR}, orbital-selective Mott phase~\cite{Anisimov01}, and excitonic magnets~\cite{Khaliullin01,Kaushal01}. In addition, the concomitant orbital and magnetic orders have also been of great interest in multi-orbital systems for many decades~\cite{Khomskii01, KhomskiiBook}. For example, the $e_{g}$ perovskite-type materials such as cubic LaMnO$_{3}$~\cite{Pavarini01,DagottoCMR} and KCuF$_{3}$~\cite{Liechtenstein01}, and quasi two-dimensional K$_2$CuF$_4$~\cite{Itol01} provide canonical examples of coexisting orbital and magnetic ordering. In all the above cases, the antiferro-orbital ordering is accompanied by ferromagnetic (FM) ordering in the ab-plane, in compliance with the GK rules~\cite{KhomskiiBook}.

Recently, altermagnetism~\cite{Hayami01,Smejkal01,Yuan01,Mazin01}, a collinear staggered magnetic state with zero net magnetization but accompanied by substantial non-relativistic spin splitting in electronic bands, has also been proposed to be present in multi-orbital systems given that co-existing staggered orbital and staggered spin order is stabilized~\cite{Leeb01}. As a proof of concept, a minimal two-orbital Hamiltonian with effective interactions including the nearest-neighbor antiferromagnetic (AFM) exchange and antiferro-orbital (AFO) exchange was used to model the emergence of spontaneous altermagnetic metal~\cite{Leeb01}. Such interactions, however, violate the second GK rule which instead mandates ferromagnetic exchange between the orthogonal occupied orbitals~\cite{KhomskiiBook}. It is therefore unclear to what extent the two-orbital model applies to real magnetic materials.

\begin{figure*}[!t]
\hspace*{-0.2cm}
\vspace*{0cm}
\includegraphics[width = 17.0cm]{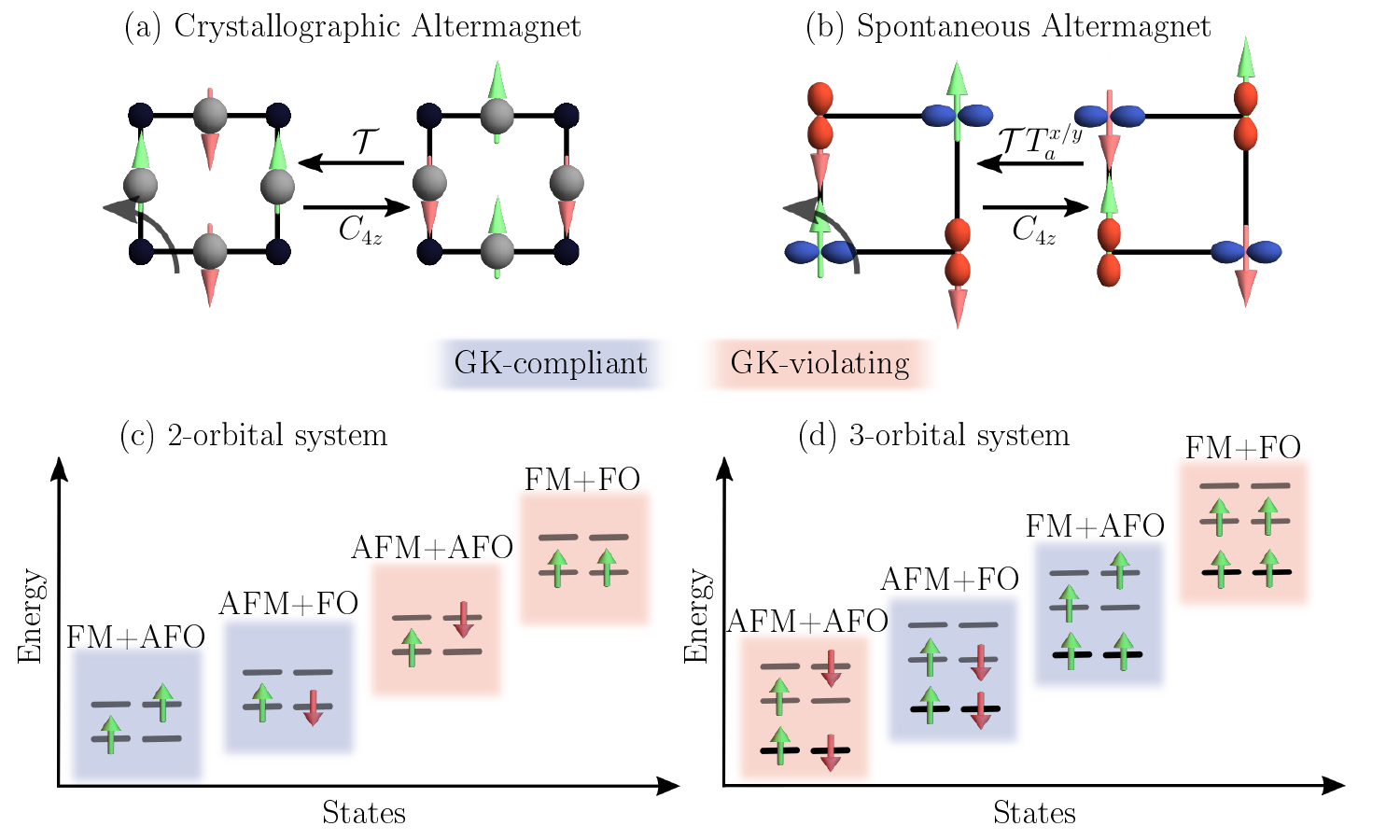}
\caption{Pictorial representation of crystallographic altermagnet and spontaneous altermagnet is depicted in panels (a) and (b), respectively. The operators $C_{4z}, {\mathcal{T}}$, and $T_{a}^{x/y}$ denote $90^ {\circ}$ rotation around the $z$-axis, time-reversal, and translation by lattice spacing along $x/y$ direction, respectively. Panel (c) shows the typical energy spectrum of a quarter-filled 2-orbital system, with GK-compliant states being most stable. The achievable spectral ordering of the 3-orbital system, specifically $t_{2g}^{2}$ under a tetragonal compressed environment, is shown in panel (d), with AFO+AFM stabilized by the additional antiferromagnetic exchange of the new half-filled third orbital.}
\label{fig0}
\end{figure*}

The crucial distinction between the more prevalent {\it crystallographic} altermagnets and the above {\it spontaneous} altermagnets is illustrated in Fig.\ \ref{fig0}. In the former case the crystallographic environment provided by the non-magnetic atoms surrounding the magnetic transition metal elements lowers the symmetry such that the antiferromagnetically correlated magnetic atom positions are related by a $C_{n}$ rotation but not by translation or inversion, see Fig.~\ref{fig0}(a). By contrast, in the spontaneous altermagnet, the crystal lattice is fully symmetric and staggered orbital and magnetic orders emerge spontaneously from Coulomb interactions, which then reduce the symmetry of the state to $C_{4}$ rotations and bring about  $d$-wave altermagnetism as depicted in Fig.~\ref{fig0}(b).

Crystallographic altermagnets have gained a lot of attention in the last few years because of their potential applications in spintronics, magnonics, and other interesting possibilities related to superconductivity and exotic topological states~\cite{YLi01,Zhu01,Monkman01,Heung01,JKrempasky01}. 
Multiple materials, including $\textrm{MnTe}$~\cite{JKrempasky01,Lee01,Osumi01}, $\textrm{CrSb}$~\cite{Ding01,Lu01,CLi01}, $\textrm{KV}_{2}\textrm{Se}_{2}\textrm{O}$~\cite{BJiang01}, and Rb$_{1-\delta}$V$_{2}$Te$_{2}$O~\cite{FZhang01}, were shown to be crystallographic altermagnets with compelling results from angle-resolved photoemission experiments and neutron scattering. 
On the other hand, as of now, there is no experimental evidence for spontaneous altermagnets, and this class of materials also remains theoretically much less explored compared to crystallographic altermagnets. Moreover, to stabilize spontaneous altermagnetism, violation of the GK rules appears unavoidable which implies an interesting open fundamental question as to their very existence.

In a typical quarter-filled two-orbital system, the GK-compliant FM+AFO and AFM+FO states are generally found to be energetically most stable. Fig.~\ref{fig0}(c) illustrates the schematic low-energy spectrum of a minimal 2-orbital system. Two sites are shown only for illustrative purposes. The FM state is favored in conjunction with an AFO order, as Hund’s coupling prefers spin-aligned intermediate states and thereby generates ferromagnetic exchange. In contrast, AFM order is favored in conjunction with FO order due to the conventional Anderson superexchange mechanism. However, the inclusion of a half-filled third orbital can modify the energy hierarchy by strengthening the net AFM exchange while maintaining intra-atomic spin alignment due to Hund’s coupling, and hence stabilize the altermagnetic (AFM+AFO) ground state. The low-energy spectrum for such a system is exhibited in Fig.~\ref{fig0}(d). 

The above intuitive picture, therefore, identifies systems with three active orbitals as promising candidates for realizing spontaneous altermagnetism. Interestingly, LaVO$_{3}$, vanadium-based cubic perovskite-type material with active $t_{2g}$ orbitals and 2 electrons per V$^{3+}$ ion, has been shown to exhibit AFO+AFM ordering in the ab-plane, believed to be driven mainly by superexchange~\cite{Miyasaka01,XJZhang01,Wohlfeld01}, long before the advent of altermagnetism. Naturally, cubic vanadates have re-gained attention in the context of altermagnetism~\cite{Cuono01}. Moreover, synthesizing the quasi-two-dimensional analogs of cubic vanadates with AFO+AFM ordering will be of interest in the context of two-dimensional altermagnets and their related technological applications. 

In the present work, we conduct a comprehensive theoretical study of two-dimensional $t_{2g}$ systems to ($i$) investigate the stability criteria for spontaneous altermagnetism and ($ii$) explore various novel aspects of this exotic state. The key findings of our work can be summarized as follows. First, using the unrestricted Hartree-Fock method, we solve the three-orbital Hubbard model on the square lattice comprising the $xy$, $xz$, and $yz$ orbitals at each site. We establish its phase diagram as a function of multiple parameters including the on-site Coulomb interaction strength $U$, tetragonal crystal field splitting $\Delta$, and the relative strength of intra-orbital hoppings. The main result of this mean-field study is a finding of the altermagnetic (ALM) Mott insulator as a ground state in a large part of the phase diagram in the compressed tetragonal crystal field environment characterized by $\Delta<0$. For the large-$U$ limit we next employ the second-order perturbation theory and derive an effective spin-orbital Hamiltonian which we solve using the density matrix renormalization group (DMRG) technique in cylindrical geometries and confirm the stability of ALM order in the presence of quantum fluctuations.  We also show, by performing classical Monte-Carlo simulations on the spin-orbital model, that the thermal fluctuations can drive the ALM state to either purely AFO or AFM phases, depending on the strength of Hund's coupling. Importantly, these transitions can be observed in resonant X-Ray and neutron diffraction experiments.

Similar to spin-splitting in single-electron excitations,  crystallographic altermagnets are also predicted to show chiral splitting in their magnon spectra~\cite{Smejkal02,Maier01,Gaitan01,Kaushal02}. This has been observed in recent inelastic neutron scattering experiments~\cite{Liu01,Singh01} and may have magnonic applications. The anisotropic Anderson superexchange, related to the reduced spatial symmetry, is the underlying reason for the chirally split magnons. In spontaneous altermagnets, the lattice symmetry is not broken explicitly and the existence of the chiral splitting in magnon spectra  remains an open question. By performing linearized time-dependent Hartree-Fock calculations, we investigate the spin excitations in the multi-orbital Hubbard model and, for the first time, establish the presence of chirally split magnons in the context of spontaneous altermagnetism. Interestingly, we find that the chiral splitting shows a $1/U^{3}$ dependence, indicating that it is driven by a spin exchange generated by fourth-order processes. The above description is further confirmed by the linear spin-orbital wave theory results on our modified spin-pseudospin model, which includes the spin-exchange terms allowed by symmetry that presumably can be derived from fourth-order perturbation theory. 

In addition to magnons, the systems with active orbital degrees of freedom are generally expected to have collective orbital excitations known as orbitons. Unlike magnons, experimentally observing orbitons has been challenging~\cite{Saitoh01,Gruninger01}, and so far, only resonant inelastic X-ray scattering-based experiments have provided convincing evidence~\cite{Schlappa01,JLi01}. We propose here a realistic and previously unexplored route to hybridize the orbitons with magnons in spontaneous altermagnets. We show that modest  external in-plane magnetic fields, in the presence of ubiquitous weak atomic spin-orbit coupling, can enable the magnon-orbiton hybridization, facilitating experimental detection of orbitons directly in inelastic neutron scattering experiments. Finally, we demonstrate the presence of a novel hybrid chiral magnon-orbiton mode possessing finite orbital polarization, and hence capable of producing orbital currents under a thermal gradient.

\section{Model and its mean-field analysis}
To investigate the effect of correlations on $t_{2g}$ electrons, we consider a standard three-orbital Hubbard Hamiltonian~\cite{DagottoCMR} on a square lattice comprising the kinetic energy $H_{K}$, and the Coulomb interaction $H_{\textrm{int}}$.
The kinetic part is given by
\begin{multline}
\hspace{0pt}
H_{K}=\sum_{\substack{{ i},\alpha,\beta, s \\ \hat{\gamma} \in\{{  {\hat  { x} } , {\hat  {y} }}\} }}t_{\alpha,\beta}^{\gamma}(c_{{i}+ \hat{\gamma} s \alpha}^{\dagger}c_{{i} s \beta} + \textrm{H.c}) + \Delta\sum_{i}  n_{i xy},
\hspace{0pt}
\end{multline}
 where $c_{i s \alpha}^{\dagger}$ denotes the electron creation operator at site ${i}$ and orbital $\alpha \in \{xy,xz,yz\}$ with spin $s$. The electron density operator is denoted by $n_{i\alpha}=\sum_{s} c_{i s \alpha}^{\dagger}c_{i s \alpha}$, and $\Delta$ is the energy of $xy$-orbital  emulating the tetragonal crystal field splitting. The $t_{\alpha,\beta}^{\gamma}$ represents the elements of the $3\times 3$ hopping matrices $t^{ {\hat {x}} ( {\hat{y}} )}$ connecting the nearest neighbor bonds along the $\hat{x} (\hat{y})$ direction. Throughout this work, we impose the following realistic conditions on the hopping matrices: ($i$) $t^{\hat {x}}$ and $t^{\hat {y}}$ are connected by the transformation $xz(yz)\rightarrow yz(xz)$, and ($ii$) the inter-orbital hoppings are set to zero. The above assumptions reduce the hopping matrices, written in orbital basis $\{ xy, xz, yz\}$, to the following form,
\begin{multline}{\label{hopmats}}
t^{ \hat{ {x} } }=\begin{pmatrix}
t_{z} & 0 & 0 \\
0 & t_{x} & 0 \\
0 & 0 & 0
\end{pmatrix}, \ \ \ 
t^{ \hat{ {y} } }=\begin{pmatrix}
t_{z} & 0 & 0 \\
0 & 0 & 0 \\
0 & 0 & t_{y}
\end{pmatrix}.
\hspace{20pt}
\end{multline}
From now on we fix $t_{z}=1.0$, keeping it as a unit of energy in all our calculations. We further enforce  $t_{x}=t_{y}$ as discussed above, and use $t_{x}$ as free parameter. 

The on-site multi-orbital Hubbard interaction reads
\begin{multline}
H_{\textrm{int}} = U\sum_{i,\alpha}n_{i\uparrow \alpha}n_{i\downarrow \alpha} + (U' - J_{H}/2)\sum_{i,\alpha < \beta}n_{i \alpha} n_{i \beta} \\
- 2J_{H}\sum_{i, \alpha < \beta}{\mathbf S}_{i\alpha}\cdot {\mathbf{S}}_{i\beta} + J_{H}\sum_{i, \alpha<\beta}(P_{i\alpha}^{\dagger}P_{i\beta} + \textrm{H.c.}),
\end{multline}
where 
\begin{multline}
{\mathbf {S}}_{i\alpha}=\frac{1}{2}\sum_{s s'} c_{i s \alpha}^{\dagger} {\bm{\sigma}}_{s s'} c_{i s' \alpha}, {\textrm{ and }} P_{i\alpha}=c_{i\downarrow \alpha}c_{i\uparrow \alpha}
\end{multline}
define the spin and pair annihilation operators, respectively, at the site $i$ and orbital $\alpha$. The intra and inter-orbital  Coulomb repulsions are parametrized by $U$ and $U'-J_{H}/2$, respectively. We employ the standard relation $U' = U - 2J_{H}$, where $J_{H}$ represents the Hund's coupling. Throughout this work, we fix the average electronic density $n$ to 2 electrons per site. 

\begin{figure}[!t]
\hspace*{-0.2cm}
\vspace*{0cm}
\includegraphics[width = 8.5cm]{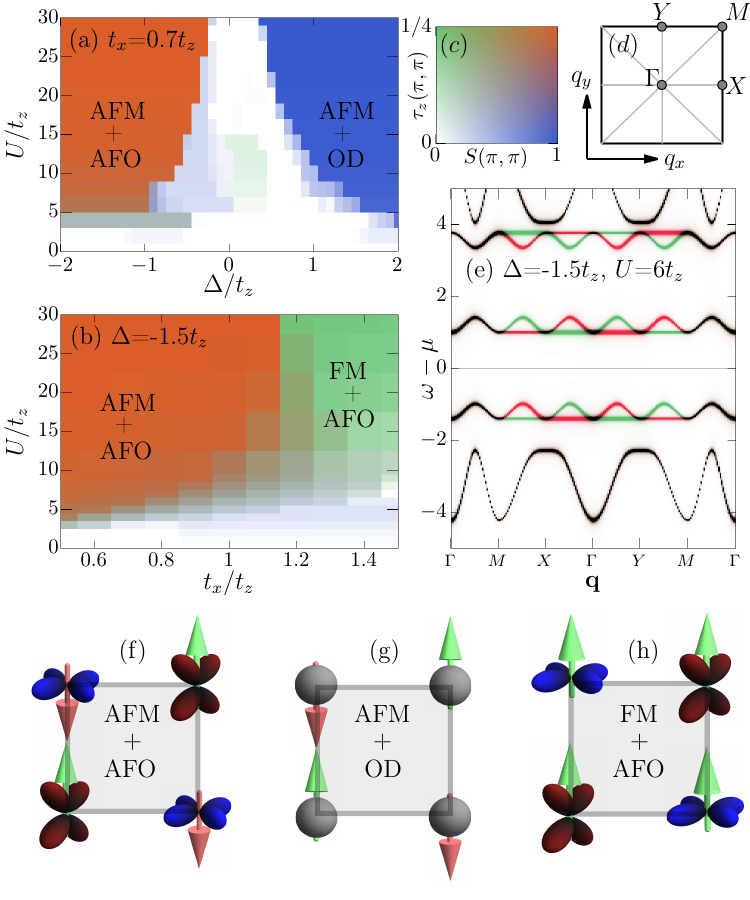}
\caption{Unrestricted Hartree-Fock results:  Panels (a) and (b) depict the $U$ vs $\Delta$ and $U$ vs $t_{x}/t_{z}$ phase diagrams, respectively. Panel (c) depicts the colors chosen in the phase diagrams, according to the $(\pi,\pi)$ structure factor values of spin and pseudospin. The full Brillouin zone is depicted in panel (d). The electronic bands exhibiting altermagnetic splitting are shown in panel (e). Panels (f), (g), and (h) show the pictorial representation of the main states we found in the phase diagrams. The abbreviations AFM, FM, AFO, and OD denote antiferromagnetic, ferromagnetic, antiferro-orbital, and orbital-disordered, respectively.}
\label{fig1}
\end{figure}

Next, we discuss the magnetic instabilities of the system within the mean-field approximation. The three-orbital Hubbard model defined above is solved using the unrestricted Hartree-Fock (HF) method, for the system size of $12\times 12$. We fix $J_{H}/U$ to a robust value of 0.2 and  investigate the effect of tuning other parameters including $U/t_{z}$, $\Delta/t_{z}$, and $t_{x}/t_{z}$ to find the stability criteria for the spontaneous altermagnetic state.  In the resulting phase diagrams Fig.~\ref{fig1}(a,b) the color corresponds to staggered magnitude of the
spin and the electronic occupation in $xz$-$yz$ orbitals, which are quantified by the structure factors at crystal momentum $(\pi,\pi)$, see the inset of Fig.~\ref{fig1}(c). The structure factors for the spin and $z$-component of pseudospin are defined as $S({\bf q})=1/(L_{x}L_{y})^{2}\sum_{ij}\langle {\mathbf S}_{i}\cdot {\mathbf S}_{j}\rangle e^{\dot{\iota} {\bf q}\cdot ({\bf r}_{i}-{\bf r}_{j})}$ and $\tau_{z}({\bf q})=1/(L_{x}L_{y})^{2}\sum_{ij}\langle {\mathbf \tau }_{iz} {\mathbf \tau }_{iz} \rangle e^{\dot{\iota} {\bf q}\cdot ({\bf r}_{i}-{\bf r}_{j})} $, respectively, where the local spin ${\bf S}_{i}=\sum_{\alpha}{\bf S_{i\alpha}}$, and ${\tau}_{iz}$ is $z$-component of the pseudospin-1/2 operator defined as
\begin{equation}\label{tauCRelation}
{\bm \tau}_{i}=\frac{1}{2}\! \!\!\!\! \sum_{\substack{s \\ \alpha, \beta \in \{yz,xz\}}}\! \! \! \!\!\!c_{is\alpha}^{\dagger}{\bm{\sigma}}_{\alpha\beta}c_{is\beta}.
\hspace{10pt}
\end{equation}

First, we discuss the relevance of tetragonal distortion in stabilizing altermagnetism. To this end we explore the $U$ vs.\ $\Delta$ phase diagram, as shown in Fig.~\ref{fig1}(a), while fixing $t_{x}/t_{z}=0.7$.
We notice that for $\Delta<0$, that is, the region with tetragonal compression along the $z$-direction, an AFM+AFO insulating state is stabilized as indicated by non-zero values of $S(\pi,\pi)$ and $\tau_{z}(\pi,\pi)$. In this phase, the $xy$-orbital is nearly half-filled, and $xz$ and $yz$ orbitals share one electron in a staggered fashion. The pictorial description of this state, only including $xz$ and $yz$ orbitals, is given in Fig.~\ref{fig1}(f). The single-particle bands for a representative parameter point in this phase are shown in Fig.~\ref{fig1}(e), depicting the $d_{x^{2} - y^{2}}$-wave nature of non-relativistic splitting in spin-up (green) and spin-down (red) electronic bands, and with a gap near the chemical potential ($\mu$) indicating an altermagnetic Mott insulating state. We remark that, recently, a variational cluster-approximation-based calculations have also obtained similar spin-split single-particle excitations for specific parameters relevant for the $ab$-plane of cubic LaVO$_3$ material~\cite{Daghofer02}, supporting our mean-field results. Moreover, considering the phase diagram we notice that Coulomb interaction strength and tetragonal compression behave cooperatively in stabilizing the ALM state, namely, the larger $|\Delta|$ helps to attain the ALM state for smaller values of $U$ and vice versa. The above result suggests that for materials with $t_{2g}^{2}$ filling and weak correlations, applying pressure along the $z$-direction to achieve stronger tetragonal compression may help to stabilize altermagnetic order. In contrast to the above, in the $\Delta>0$ region, we find an AFM insulator without any orbital ordering, denoted as AFM + OD (orbital-disordered), schematically depicted in Fig.~\ref{fig1}(g). In this region, which mimics the crystal field splitting of a tetragonally elongated environment, the higher energy $xy$ orbital is almost empty, and both $xz$ and $yz$ orbitals are nearly half-filled, driving the AFM exchange.

 So far, we have numerically established that the condition $\Delta<0$ is crucial to achieving the  ALM order in the $t_{2g}^{2}$ system. Now we discuss the effect of the relative strengths of intra-orbital hopping amplitudes controlled by $t_{x}/t_{z}$, for which we explored the $U$ vs.\ $t_{x}$ phase diagram in Fig.~\ref{fig1}(b), fixing $\Delta=-1.5$. We found that increasing $t_{x}$ destabilizes the ALM state towards a state with FM and AFO ordering, depicted in Fig.~\ref{fig1}(h), which is compatible with the GK rules. This result indicates that robust $xy-xy$ orbital hopping is crucial to violation of the GK rules.
 
 In summary, our zero-temperature mean-field results suggest the presence of an ALM Mott insulator for intermediate to strong Coulomb correlation values, given a tetragonally compressed crystal field environment and robust intra-orbital hopping amplitudes in $xy$ orbital relative to $xz$ and $yz$ orbitals.

\section{Large-U limit models}
\subsection{Effective spin-orbital model}
As revealed in our Hartree-Fock study, the strong Coulomb correlations play an important role in stabilizing the altermagnetic Mott insulator. In this section, we discuss the effective spin-orbital models which we will use to further investigate the strongly correlated altermagnetic insulator and the related properties. In the large-$U$ limit, the Hund's interaction favors on-site ferromagnetic coupling, leading to the formation of local $S=1$ moments. Moreover, in $t_{2g}^{2}$ Hilbert space, the local orbital angular momentum $L$ is unquenched. The effective angular momentum is defined as
\begin{equation}
{L}_{i\eta}=-\dot{\iota}  \sum_{s\alpha\beta}c_{is\alpha}^{\dagger}c_{is\beta}\epsilon_{\eta \alpha \beta},
\end{equation}
where $\eta \in \{x,y,z\}$ and the orbitals $\{ \alpha,\beta\} \in \{xy \equiv z, xz  \equiv y, yz  \equiv x\}$. In the large-$U$ limit, we expect $\langle L_{i}^{2}\rangle \approx 2$, as also confirmed by our exact diagonalization calculations, see SM~\cite{SM}. These considerations suggest that spin-orbital model within the local $S=L=1$ Hilbert space should capture the low-energy physics. Performing second-order perturbation theory in the large-$U$ limit using the hopping matrices defined in Eq.~\eqref{hopmats}, we obtain the effective spin-orbital ($S$-$L$) Hamiltonian
\begin{equation}\label{HSLeff}
H^{SL} = \Delta \sum_{i}L_{iz}^{2} + \sum_{ \langle {i} {j}\rangle }(H^{SQ}_{{ i},{j}} + H^{SL}_{{ i},{ j}} + H^{Q}_{{i},{j}} + H^{L}_{{i},{j}})
\end{equation}
where the first term corresponds to the tetragonal crystal field splitting and the rest describes the nearest-neighbor exchange terms. Assuming ${j}={i}+{\gamma}$, where $\gamma\in \{x,y\}$ denotes the bond direction, the four exchange terms are given as
\begin{multline}\label{HSQ}
H^{SQ}_{{i},{j}} = {\bf{S}}_{{i}}\cdot {\bf{S}}_{{j}}( f^{SQ}_{1}(t_{\Bar{\gamma}}^{2}L_{{ i}\Bar{\gamma}}^{2}L_{{j} \Bar{\gamma}}^{2} + t_{z}^{2}L_{{ i}z}^{2}L_{{j}z}^{2})
\hspace{100pt}
\\
+ f^{SQ}_{2}( t_{\Bar{\gamma}}^{2}(L_{{i}\Bar{\gamma}}^{2}+L_{{j}\Bar{\gamma}}^{2}) + 
t_{z}^{2}(L_{{i}{z}}^{2}+L_{{j}{z}}^{2}) )
\\
+f_{3}^{SQ}t_{\Bar{\gamma}}t_{z}Q_{{ i}}^{\Bar{\gamma}z}Q_{{ j}}^{\Bar{\gamma}z})
\hspace{10pt}
\end{multline}
\begin{multline}\label{HSL}
H^{SL}_{{i},{j}}= {\bf{S}}_{{i}}\cdot {\bf{S}}_{{j}} (f^{SL} t_{\Bar{\gamma}}t_{z}  L_{{i}\gamma}L_{{j}\gamma})
\hspace{85pt}
\end{multline}
\begin{multline}\label{HQ}
H^{Q}_{{i},{j}} = f^{Q}_{1}(t_{\Bar{\gamma}}^{2}L_{{i}\Bar{\gamma}}^{2}L_{{j} \Bar{\gamma}}^{2} + t_{z}^{2}L_{{i}z}^{2}L_{{j}z}^{2}) +f_{3}^{Q}t_{\Bar{\gamma}}t_{z}Q_{{i}}^{\Bar{\gamma}z}Q_{{j}}^{\Bar{\gamma}z},
\hspace{100pt}
\\
+ f^{Q}_{2}( t_{\Bar{\gamma}}^{2}(L_{{i}\Bar{\gamma}}^{2}+L_{{j}\Bar{\gamma}}^{2}) + 
t_{z}^{2}(L_{{i}{z}}^{2}+L_{{j}{z}}^{2}) ),
\hspace{10pt}
\end{multline}
\begin{multline}\label{HL}
H^{L}_{{i},{j}}= f^{L} t_{\Bar{\gamma}}t_{z}  L_{{i}\gamma} L_{{j}\gamma}.
\hspace{125pt}
\end{multline}

The terms $H_{ij}^{SQ}$ and $H_{ij}^{SL}$ capture the interactions between the spin dipole and orbital quadrupole moments, and the spin dipole and orbital dipole moments, respectively. The next two terms $H_{ij}^{Q}$ and $H_{ij}^{L}$ represent the interactions within the orbital space, namely the quadrupole-quadrupole and the dipole-dipole interactions, respectively.  The quadrupole operator $Q_{i}^{\bar{\gamma} z}=L_{i\bar{\gamma}}L_{iz}+L_{iz}L_{i\bar{\gamma}}$, and $\bar{\gamma}=x(y)$ for $\gamma=y(x)$. The exchange parameters $f_{1}^{SQ}, f_{2}^{SQ}, f_{3}^{SQ}, f^{SL}, f_{1}^{Q}, f_{3}^{Q}, f_{2}^{Q}$, and $f^{L}$ are obtained within the second-order perturbation theory as functions of $U$ and $J_{H}/U$; details are given in SM~\cite{SM}. We confirmed that the effective Hamiltonian Eq.\ \eqref{HSLeff} faithfully represents the spin-orbital physics of the multi-orbital Hubbard model defined in Sec. III by comparing the exact low-energy spectrum of both models on a small 2-site cluster, for various $\Delta$ and $J_{H}/U$, in the large-$U$ limit~\cite{SM}. 

\begin{figure}[!t]
\hspace*{-0.2cm}
\vspace*{0cm}
\includegraphics[width = 8.5cm]{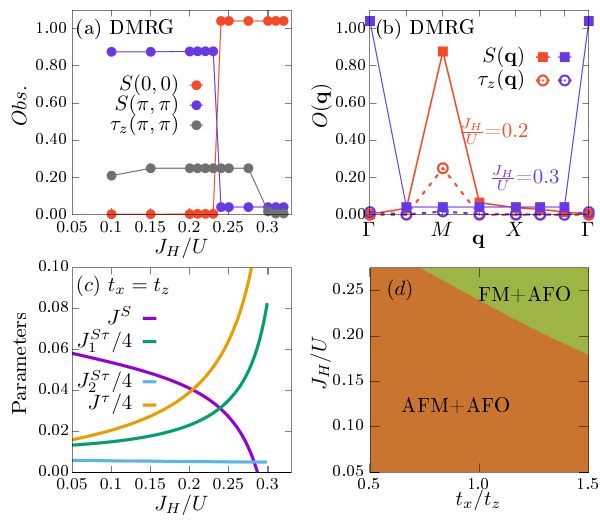}
\caption{Panels (a) and (b) depict the DMRG results for system size of $8\times4$ with cylindrical boundary conditions. Panel (c) shows the exchange parameters dependence on $J_{H}/U$, for $S$-$\tau$ model. The $J_{H}/U$ vs $t_{x}/t_{z}$ phase diagram, illustrated in panel (d), is constructed by comparing energies of AFM+AFO, AFM+FO, FM+AFO, and FM+FO states. We fix the parameters $(\Delta,t_{x},U)=(-1.5,1.0,20)t_{z}$ for the panels (a,b,c).}
\label{fig2}
\end{figure}

{\subsection{Stability of ALM state to quantum fluctuations}} 

We used the effective $S$-$L$ model to confirm the stability of the AFM+AFO state against the competing magnetic instabilities beyond the mean-field treatment discussed Sec.\ III. Specifically, we studied the effect of quantum fluctuations by performing DMRG calculations employing the ITensor library~\cite{Fishman01} with the bond dimension up to 2500 to ensure accurate results. Using the guidance of our mean-field theory results, we fix the parameters $\Delta=-1.5$, $t_{x}=1.0$, and $U=20$ so as to maintain the large-$U$ limit. Smaller  $t_{x}$ values were also studied~\cite{SM}, essentially led to similar conclusions described below. We solved the $(L_{x},L_{y})=(8,4)$ lattice with periodic (open) boundary conditions along the $x$ $(y)$ direction, realizing the cylindrical geometry.  In Fig.~\ref{fig2}(a), we show the evolution of $S(\pi,\pi)$, $S(0,0)$, and $\tau_{z}(\pi,\pi)$ varying $J_{H}/U$. The structure factors for $J_{H}/U=0.2$, and $0.3$ are shown in Fig.~\ref{fig2}(b). Importantly, for a wide range of weak to modest values of $J_{H}/U$,  we found that the AFM+AFO state is stable unless Hund's coupling is very strong in which case the FM+AFO or FM+OD states are instead realized. The stability of the AFM+AFO state beyond our mean field theory results is promising, as quantum fluctuations tend to be significant in quasi-2D materials.

~

{\subsection{Effective spin-pseudospin model}}

For $\Delta<0$, when the $xy$ orbital is nearly half-filled, we can 
approximate $L_{z}^{2}\rightarrow 1$, $Q^{\alpha z}\rightarrow 0$ , and define $\tau_{z}=(L_{x}^{2}-L_{y}^{2})/2$, see SM~\cite{SM}. Using~\cref{HSLeff,HSQ,HSL,HQ,HL}, we can readily derive the following Kugel-Khomskii type spin-pseudospin ($S$-$\tau$) Hamiltonian
\begin{multline}\label{HSTau}
H^{S\tau} =  \sum_{\langle {{ij}} \rangle} {{\bf S}_{i}}\cdot  {{\bf S}_{j}} (J^{S} + J_{1}^{S\tau}\tau_{ {i} z}\tau_{ {j} z}   \pm J_{2}^{S\tau} (\tau_{ { i} z} + \tau_{ {j} z}) )
\hspace{100pt}
\\
+ J^{\tau} {\tau}_{ {i}z } {\tau}_{ {j}z}.
\hspace{120pt}
\end{multline}
The exchange parameters $J^{S}$, $J_{1}^{S\tau}$, $J_{2}^{S\tau}$, and $J^{\tau}$ can be represented as functions of $t_{x/y}$, $t_{z}$, $U$, and $J_{H}/U$ ~\cite{SM}. The $+(-)J_{2}^{S\tau}$ term corresponds to the nearest neighbor bonds in $x(y)$ direction. All the exchange parameters are positive for any realistic $J_{H}/U$ value, as shown in Fig~\ref{fig2}(c). The orbital  ($J^{\tau}$) and spin ($J^{S}$) exchanges favor the AFO and AFM states, respectively, while the spin-orbital interaction represented by $J_{1}^{S\tau}>0$ would prefer FM+AFO or AFM+FO states that comply with the GK rules.

To develop an intuitive understanding of the stability of the AFM+AFO state  we now give simple energetic arguments based on Eq.~\eqref{HSTau}. For this purpose, we work in the classical limit and consider only four collinear states: FM+FO, FM+AFO, AFM+FM, and AFM+AFO. The FM+FO state is the highest in energy as it is incompatible with all exchange parameters. 
Importantly, it is guaranteed that the AFM+AFO state is stable against the FM+AFO and AFM+FO states, given $J^{S}>J_{1}^{S\tau}/4$ and $J^{\tau}>J_{1}^{S\tau}$, respectively. We notice that increasing $J_{H}/U$ reduces $J^{S}$ and enhances both $J^{\tau}$ and $J^{S\tau}_{1}$, see Fig.~\ref{fig2}(c), eventually leading to $J^{S}<J^{S\tau}_{1}/4$ which is detrimental to the AFM+AFO state. In this large Hund's coupling limit, the antiferro-orbital ordering driven by large $J^{\tau}$ provides effectively a ferromagnetic spin exchange, via $J^{S\tau}_{1}$, stabilizing the FM+AFO state. We found that a similar AFM+AFO to FM+AFO transition is driven by increasing $t_{x}/t_{z}$, as depicted in the $J_{H}/U$ versus $t_{x}/t_{z}$ phase diagram in Fig.~\ref{fig2}(d), which is constructed just by comparing the energy of the above mentioned classical ansatz states. The above $t_{x}/t_{z}$ dependence can be understood from the dominant contribution of the Anderson superexchange generated from the half-filled $xy$ orbital described by the $t_{z}^{2}/U$ term in 
\begin{equation}
J^{S}= \frac{t_{x/y}^{2}}{U}\left(-\frac{\zeta}{2(1-3\zeta)} + \frac{1+\zeta}{4(1+2\zeta)}\right)
+ \frac{t_{z}^{2}}{U}\frac{1+\zeta}{(1+2\zeta)},
\end{equation}
where $\zeta=J_{H}/U$. The $t_{z}^{2}/U$ term plays a crucial role in stabilizing the AFM state, as the $t_{x/y}^{2}/U$ term is much weaker in magnitude because of the oppositely-signed internal terms.

In summary, the above discussion again demonstrates that the contribution of the half-filled third orbital (the $xy$ orbital in our case), which does not participate in the staggered orbital ordering, to the net antiferromagnetic superexchange $J^{S}$ is an important ingredient stabilizing the GK–violating AFM+AFO state. In the absence of sufficiently strong $t_{z}$, the GK-compliant FM+AFO state is stabilized, as seen in Figs.~\ref{fig1}(b) and \ref{fig2}(d). The above enhancement mechanism of antiferromagnetic exchange is absent in canonical two-orbital systems, and therefore, such systems naturally comply with the GK rules.

~

\section{Hund's role in determining the temperature hierarchy}

In the preceding sections, we have established the AFM+AFO state as the ground state in both the mean field and DMRG studies. In the present section we employ the $S$-$L$ model to study the phase transitions driven by the thermal fluctuations. To proceed we make a semiclassical approximation and describe our system as an ensemble of entanglement-free direct product states $\prod_{i}|\Psi_{i}^{S}\rangle|\Psi_{i}^{L}\rangle$. This approximation should be reasonable given that our DMRG results found a robust AFO+AFM state, suggesting only a mild effect of quantum fluctuations, which we expect to be further suppressed at finite temperatures.

\begin{figure}[!t]
\hspace*{-0.2cm}
\vspace*{0cm}
\includegraphics[width = 8.5cm]{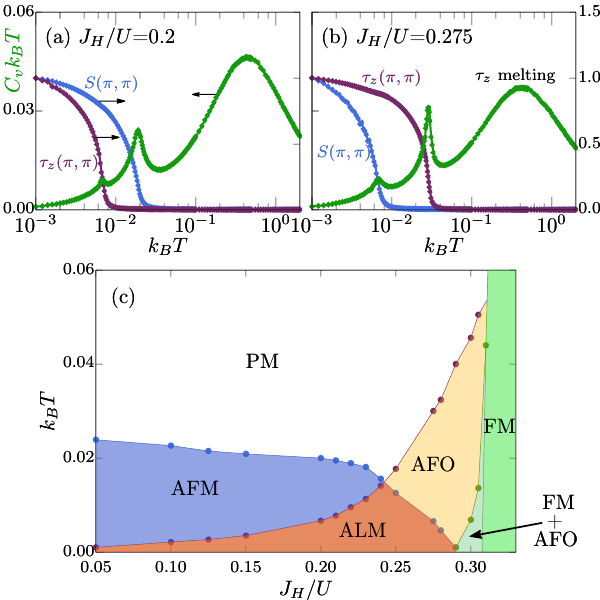}
\caption{The panels (a,b) show the specific heat, spin and pseudospin $(\pi,\pi)$ structure factors evolution with temperature, depicting the phase transitions. The temperature vs $J_{H}/U$ phase diagram is shown in panel (c). The parameters $(\Delta,t_{x},U)=(-2, 0.7,20)t_{z}$ are fixed for all the above panels.}
\label{fig3}
\end{figure}

To faithfully represent the local $S=1$ and $L=1$ states, we employ the  ${\mathbb{CP}}^{2}$ representation~\cite{SM} and perform classical Monte Carlo on the $S$-$L$ model. We calculate multiple observables, including the specific heat $C_{v}$, spin and orbital structure factors, while spanning the temperature $T$ for various Hund's coupling values, for details, see~\cite{SM}. The $\Delta=-2.0t_{z}$, and the exchange parameters corresponding to $(t_{x}, U)$=$(0.7,20)t_{z}$ are used. We show $C_{v}k_{B}T$, $S(\pi,\pi)$, and $\tau_{z}(\pi,\pi)$, as a function of $T$ for the $J_{H}/U$ values of 0.2 and 0.275, for the system size of $12\times 12$, in Fig.~\ref{fig3}(a,b). The $S(\pi,\pi)$, and $\tau_{z}(\pi,\pi)$ are normalized by their lowest temperature value for clarity. We note that $C_{v}k_{B}T$ exhibits a 3-peak structure: The broad peak present at the highest temperature in both Fig.~\ref{fig3}(a,b) near $k_{B}T=0.3$, corresponds to the melting of the pseudospin-1/2 as large thermal fluctuations equally populate all three orbitals. For $J_{H}/U=0.2$ Fig.~\ref{fig3}(a) shows a middle peak near $k_{B}T=0.02$ which corresponds to the characteristic temperature $T_{SO}^{c}$, below which the antiferromagnetic spin correlations start to develop as indicated by the growth of $S(\pi,\pi)$ around the same temperature. The smallest temperature peak is associated with the onset of AFO ordering, marked by the development of $\tau_{z}(\pi,\pi)$ that breaks the Ising symmetry in the pseudospin sector, at the critical temperature of $T_{OO}\approx 0.008/k_{B}$. In contrast to the above, we notice that for the larger Hund's coupling of $J_{H}/U=0.275$, see Fig.~\ref{fig3}(b), the AFM and AFO transitions are present in the reverse order. The above highlights the crucial role of Hund’s coupling in determining the sequence of phase transitions upon changing the temperature.

Motivated by the above, we performed a detailed analysis for various values of $J_{H}$ to construct the temperature versus $J_{H}/U$ phase diagram as shown in Fig.~\ref{fig3}(c). We note that the critical temperatures associated with spin and orbital ordering vary strongly with $J_{H}/U$ in opposite directions, thereby producing a dome-shaped altermagnetic regime with the maximum critical temperature for ALM when $T_{SO}^{c}$ and $T_{OO}$ coincide. We also note that the system shows the FM tendencies only for the strong Hund's coupling, which is consistent with the results discussed in the previous section. The maximum critical temperature of altermagnetism $T_{ALM}$, for the parameters $(t_{x},U)$=$(0.7,20)$,  is around $0.018/k_{B}$ with $t_{z}$ being the unit of energy. In real materials, typical hopping amplitudes are 0.1-0.3eV, suggesting the $T_{ALM}$ of about 20-60K. Moreover, for the relatively modest and realistic Coulomb correlation values of $U/t_{z}$ around 8-10, with a reasonable assumption of applicability of the $S$-$L$ model, the larger $T_{ALM}$ up to 120-150K can be easily achieved. Importantly, it should be noted that the above transitions present at observable temperatures are purely driven by the Coulomb correlations i.e. no Jahn-Teller effect and lattice distortions are required. 

It should be noted that, in the two-dimensional $S$-$L$ model used for the above analysis, the true long range spin-order is excluded by Mermin-Wagner. The $T_{SO}^{c}$ corresponds to temperature below which spin-spin correlation length increases exponentially fast, and hence in our finite systems we notice effectively a long rang order. However, in any real material, the $SU(2)$-symmetry in spin is mostly explicitly broken. For example, even a small but inevitable atomic spin-orbit coupling in the presence of tetragonal crystal field splitting leads to Ising anisotropy in spin, which we also discuss in detail in the next section, and that can render the true long-range spin order in two-dimensional systems even at finite temperatures.

\section{Chiral spin-split magnons}
In this section, we investigate the spin excitations of the spontaneous multi-orbital altermagnet. Unlike the crystallographic altermagnets where the chiral splitting in magnons and related magnon-based spintronic applications have been extensively studied~\cite{QCui01}, the spin excitations of spontaneous altermagnets have not been discussed. To address this issue we present in this Section results on  spin excitations of the multi-orbital Hubbard model at zero temperature. We also construct and  study a modified spin-pseudospin model, which correctly captures the collective excitations of the model and can be used to further probe the physics of collective modes in multi-orbital altermagnets. Finally, we discuss the effect of weak atomic spin-orbit coupling on the altermagnet and its spin excitations.

\subsection{Linearized time-dependent Hartree-Fock results}

\begin{figure}[!t]
\hspace*{-0.2cm}
\vspace*{0cm}
\includegraphics[width = 8.5cm]{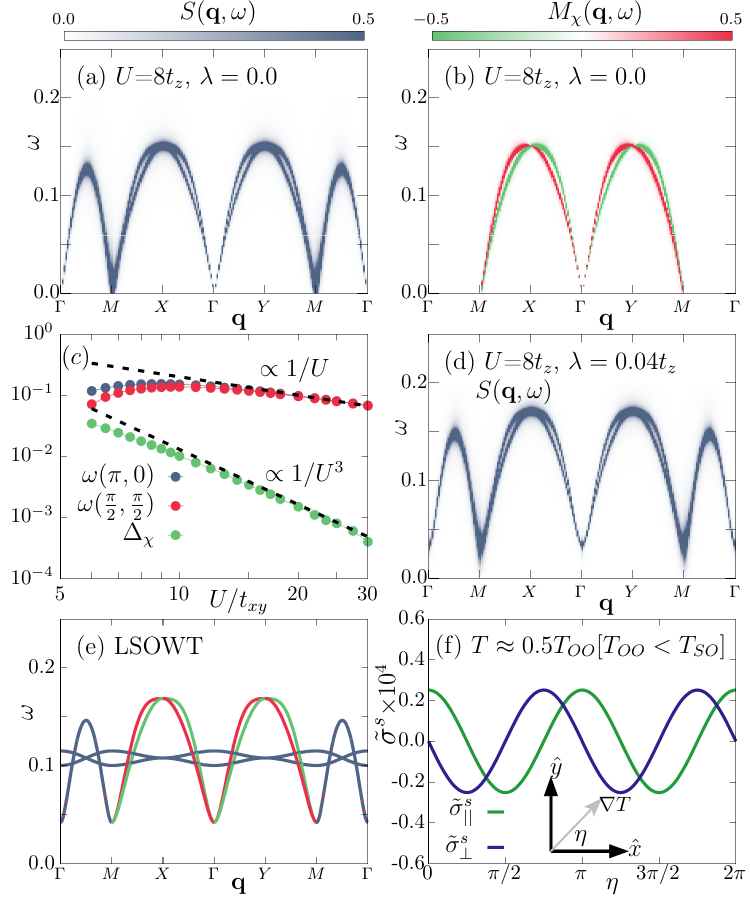}
\caption{Panels (a) and (b) display the $S({\bf q}, \omega)$ and $M_{\chi}({\bf q},\omega)$, respectively, calculated using RPA, for parameters $t_{x}$=$0.7t_{z}$, $U$=$8.0t_{z}$, $J_{H}/U$=$0.2$, $\lambda$=$0.0$ and system size of $64\times 64$. The $U$-dependence of magnon energy and chiral splitting is depicted in panel (c). Panel (d) shows the $S({\bf q}, \omega)$ for the same parameters as in panel (a), but with finite $\lambda=0.04t_{z}$. The excitation spectrum calculated using LSOWT is depicted in panel (e), for parameters $J^{S}$=0.052, $J_{1}^{S\tau}$=0.054, $J^{\tau}$=0.108, $J_{\chi}$=0.005, $K$=0.0056, and  $\lambda$=0.04. Panel (f) depicts the spin conductivities as a function of angle $\eta$.}
\label{fig4}
\end{figure}

We calculated the dynamical spin structure factor,
$S({\bf q},\omega)=\sum_{\alpha \in \{x,y,z\}} S_{\alpha\alpha}({\bf q}, \omega)$
using the linearized time-dependent Hartree-Fock approach~\cite{Giuliani01} that is equivalent to random phase approximation (RPA); for details, see SM~\cite{SM}. In Fig.~\ref{fig4}(a), we show $S({\bf q},\omega)$ for the representative set of parameters $(U, J_{H}/U, t_{x})=(8.0t_{z},0.20,0.7t_{z})$ having an altermagnetic ground state. The gapless Goldstone mode emerges from the ${\bf q}=M$ point, and its splitting into two branches is evident for the momentum points along the $M-X-\Gamma$ and $M-Y-\Gamma$ paths, whereas no splitting is present along the direct $\Gamma-M$ path. To measure the chirality of the split modes, we calculate the difference between transverse spin excitations, namely $M_{\chi}({\bf q}, \omega)=S_{+-}({\bf q},\omega)-S_{-+}({\bf q},\omega)$, while choosing the Neel vector of the ground state along $z$-direction. We notice that $M_{\chi}({\bf q}, \omega)$ depicts a clear sign reversal between the two magnon branches, and also satisfies $M_{\chi}(C_{4}{\bf q},\omega)$=$-M_{\chi}({\bf q}, \omega)$ indicating the d-wave type chiral splitting in the magnons. We verified that the chiral splitting in spin excitations is present for various values of $U$. In Fig.~\ref{fig4}(c), we depict $U$-dependence of magnon energy, estimated from $S({\bf q},\omega)$, for ${\bf q}$=$X$ and $M/2$. The magnitude of the chiral splitting $\Delta_{\chi}$, which we quantify by calculating the difference between the positions of two poles present at ${\bf q}=(\pi,\pi/2)$, is also shown in Fig.~\ref{fig4}(c). In the large-$U$ limit, magnon energy values for both ${\bf q}$=$M/2$ and $X$ decrease as $1/U$, which is expected since the spin-exchange is generated by a second-order process in hopping. In contrast, interestingly, $\Delta_{\chi}$ shows $1/U^{3}$ scaling, suggesting a fourth-order process as the underlying cause of chiral splitting. Moreover, the lower magnon energy at ${\bf q}=M/2$ than $X(Y)$ momentum points can be attributed to higher-order ring-exchange terms, following the interpretation of similar behavior observed in undoped antiferromagnetic cuprates~\cite{RColdea01}. Notably, many real materials generally lie in the intermediate Coulomb correlation regime, so moderate chiral splitting in magnons can be observed in inelastic neutron scattering experiments on $t_{2g}$ based altermagnetic Mott insulators. 

{\subsection{Effect of weak atomic spin-orbit coupling }}

\begin{figure}[!t]
\hspace*{-0.2cm}
\vspace*{0cm}
\includegraphics[width = 8.5cm]{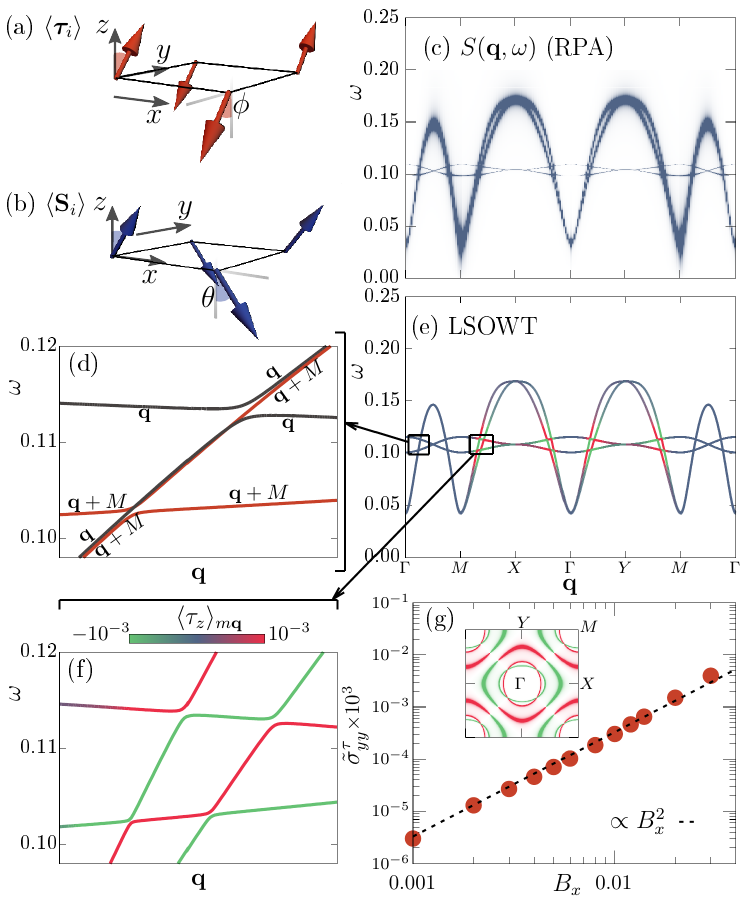}
\caption{Panel (a) depicts the canted pseudospins in the presence of finite SOC. The canted spin orientations due to finite magnetic field along spatial $x$ direction are illustrated in panel (b). Panel (c) display the spin dynamical structure factor calculated using RPA. The LSOWT results are shown in panel (e), with zoomed-in views shown in panels (d) and (f). $B_{x}=0.01$ is used for panels (d,e,f) for clearer depiction. Panel (g) shows the longitudnal orbital conductivity as a function of $B_{x}$ at $T \approx 0.5T_{OO}$.}
\label{fig5}
\end{figure}

In 3d transition metal-based materials small but non-zero atomic spin-orbit coupling (SOC) is always present. To investigate the effect of weak atomic SOC on altermagnetism and magnetic excitations, we performed the HF+RPA calculations with the following term added to the multi-orbital Hubbard Hamiltonian,
\begin{multline}\label{SOC}
H_{\textrm{SOC}} = \lambda \sum_{\substack{{i,\alpha, \alpha'}\\{\sigma, \sigma'}}}\langle \alpha|{\bf L}_{i}|\alpha '\rangle \cdot \langle \sigma |{\bf S}_{i}| \sigma' \rangle c_{i\sigma\alpha}^{\dagger} c_{i\sigma'\alpha'}.
\hspace{20pt}
\end{multline}
In our self-consistent Hartree-Fock results with small $\lambda$ we confirmed the AFM+AFO ground state. We find, in the presence of finite SOC, the local spins are aligned along the $\pm z$ Ising direction. In contrast, the local pseudospins ${\bm {\tau}}_{i}$ are rotated by a small angle of $\phi$ away from the Ising $z$-axis towards the $+(-)$ $y$-axis for the sublattice with spin along $+(-)$ $z$-direction, while still maintaining the collinear AFO order, see Fig.~\ref{fig5}(a). Notably, the above Ising anisotropy in spins and canting in pseudospins can be explained intuitively within the large-$U$ spin-pseudospin model as described later in more detail. The calculated $S({\bf q},\omega)$ for $\lambda=0.04t_{z}$ is shown in Fig.~\ref{fig4}(d), depicting essentially the same features as in Fig.~\ref{fig4}(a), except for the presence of the spin gap at ${\bf q}=M$ and $X$ points. The presence of gap in spin excitations stabilizes the altermagnetic phase at finite temperatures in two-dimensional materials.

{\subsection{Linear spin-orbital wave theory}}

The above results, in addition to establishing the presence of chiral spin-split magnons under realistic conditions for $t_{2g}$  altermagnets, also provide guidance to incorporating higher-order effects in the large-$U$ model. To probe the magnons in the large-$U$ limit, we focus on the spin-pseudospin ($S$-$\tau$) model from Eq.~\eqref{HSTau} and employ the $SU(3)$ and $SU(2)$ Schwinger boson representations for a faithful description of $S=1$ and $\tau=1/2$ spaces, respectively. We performed the linear spin-orbital wave theory (LSOWT) to describe the collective excitations, see SM for details~\cite{SM}. We note that directly using the $S$-$\tau$ model dedfined by Eq.~\eqref{HSTau} does not capture the chiral splitting in magnons, which is also consistent with the $1/U^{3}$ dependence of $\Delta_{\chi}$ in RPA results. In principle the fourth-order perturbation theory would be required to obtain the relevant terms in $S$-$\tau$ model to capture the chiral splitting. Instead of performing the cumbersome fourth-order perturbation theory, we employ symmetry analysis to identify the allowed terms that qualitatively capture the results obtained from RPA calculations. In the  $\lambda=0$ limit, any higher-order terms must satisfy the following symmetry criteria: ($i$) $SU(2)$ symmetry in spin space, ($ii$)  invariance under $C_{4v}$ point group operations of the square lattice, ($iii$) respect the time-reversal operation $\mathcal{T}$, and ($iv$) respect translational symmetry. Under time-reversal or point group operations, the  pseudospin transformation rules are obtained by using Eq.~\eqref{tauCRelation} and the response of $xz$ and $yz$ orbitals under the same transformation. The relevant transformations are summarized in Table~\ref{table1}. 

\begin{table}[h]
\caption{The action of symmetry on electron creation operators in $xz/yz$ orbitals and pseudospins ${\bm \tau}_{i}$  time reversal operator and $C_{4v}$ point group elements (except identity). Indices $i$ and $j$ represent the sites satisfying $j=\hat{\Theta}(i)$, $s$ denotes the spin, and $\eta_{s}=+1(-1)$ for $s=\uparrow(\downarrow)$.}
\label{table1}
\centering
\begin{tabular}{@{}p{2.5cm}@{\hspace{1pt}}p{3.0cm}@{\hspace{20pt}}p{2cm}}
\hline
\toprule
{Operations} & {Orbitals} & {Pseudospin} \\
$\hat{\Theta}$ & $\hat{\Theta}({{ c}}_{isxz}^{\dagger}, {{ c}}_{isyz}^{\dagger})\hat{\Theta}^{-1}$ & $\hat{\Theta}{{\bm \tau}}_{i}\hat{\Theta}^{-1}$\\ 
\hline
${\mathcal{T}}$ & $(\eta_{s}{{c}}_{i{\bar{s}}xz}^{\dagger}, \eta_{s}{{ c}}_{i{\bar{s}}yz}^{\dagger})$ &  $( \tau_{ix},-\tau_{iy},\tau_{iz})$ 
\\
$C_{4z}$ & $(-{c}_{jsyz}^{\dagger}, {{ c}}_{jsxz}^{\dagger})$ & $(-\tau_{jx},\tau_{jy},-\tau_{jz})$
\\
$C_{4z}^{-1}$ & $({c}_{jsyz}^{\dagger}, -{{ c}}_{jsxz}^{\dagger})$ & $(-\tau_{jx},\tau_{jy},-\tau_{jz})$
\\
$C_{2z}$ & $-({c}_{jsxz}^{\dagger}, {{c}}_{jsyz}^{\dagger})$ & $(\tau_{jx},\tau_{jy},\tau_{jz})$
\\
$m_{xy}$ & $({{ c}}_{jsyz}^{\dagger}, {{ c}}_{jsxz}^{\dagger})$ & $(\tau_{jx},-\tau_{jy},-\tau_{jz})$
\\
$m_{x\bar{y}}$ & $-({{c}}_{jsyz}^{\dagger}, {{ c}}_{jsxz}^{\dagger})$ & $(\tau_{jx},-\tau_{jy},-\tau_{jz})$
\\
$m_{x}$ & $(-{{ c}}_{jsxz}^{\dagger}, {{ c}}_{jsyz}^{\dagger})$ & $(-\tau_{jx},-\tau_{jy},\tau_{jz})$
\\
$m_{y}$ & $({{c}}_{jsxz}^{\dagger}, -{{ c}}_{jsyz}^{\dagger})$ & $(-\tau_{jx},-\tau_{jy},\tau_{jz})$
\\
\hline
\end{tabular}
\label{tab:3col}
\end{table}
We note that $\tau_{iz}$ is even under $\mathcal{T}$, unlike ${\bf S}_{i}$ which is odd. Under the $C_{4z}$ rotation and the mirror operation $m_{x}$, which are the generators of $C_{4v}$ group, $\tau_{z}$ is odd and even, respectively. Using the above information, we find the following third-nearest-neighbor interaction term that respects the symmetry criteria discussed before
\begin{multline}\label{ChiralTerm}
H_{\chi}=J_{\chi}\sum_{ \substack{{i},  {\gamma \in x,y} } }s_{\gamma}{\bf S}_{i}\cdot {\bf S}_{{i}+2\hat{\gamma}}(\tau_{iz}{} + \tau_{{ i}+\hat{\gamma}z}^{} + \tau_{{i}+2\hat{\gamma}z}^{}),
\hspace{0pt}
\end{multline}
where $(s_{x},s_{y})$=$(1,-1)$. Moreover, we found that similar terms including second-nearest neighbors are not allowed by mirror symmetries $m_{xy}$ and $m_{x\bar{y}}$, where  $m_{xy(\bar{y})}$ is the mirror operation from the plane perpendicular to the line $x=y(-y)$.
We note that two more third-neighbor terms like the one above can be written, which include either $\tau_{iz}\tau_{i+\gamma z}\tau_{i + 2\gamma z}$ or just $\tau_{i +\gamma z}$ instead of $\tau_{iz}{} + \tau_{{ i}+\hat{\gamma}z}^{} + \tau_{{i}+2\hat{\gamma}z}^{}$. However, all these possibilities produce a similar effect in LSOWT, so we just use Eq.~\eqref{ChiralTerm} to represent the cumulative effect. We confirm that $H_\chi$ defined in Eq.\ \eqref{ChiralTerm} is sufficient to achieve the chiral spin-split magnons. However, for a complete qualitative match, we need to incorporate a ring-exchange term of the form
\begin{multline}
H_{R} = K\sum_{\bf{ijkl}\in \square}({\bf S_{i}}\cdot{\bf S_{j}})({\bf S_{k}}\cdot{\bf S_{l}}) + ({\bf S_{i}}\cdot{\bf S_{l}})({\bf S_{j}}\cdot{\bf S_{k}})
\\ - ({\bf S_{i}}\cdot{\bf S_{k}})({\bf S_{j}}\cdot{\bf S_{l}}), 
\hspace{70pt}
\end{multline}
where the sites $i,j,k$, and $l$ are ordered clockwise on  square plaquettes. The ring exchange term is required to soften the magnon energy at ${\bf q}=M/2$  and to maintain the maxima at ${\bf q}=X (Y)$ points in the presence of chiral splitting. 
We also included the SOC in $S$-$\tau$ model, represented as $H_{\textrm{SOC}}=-\lambda\sum_{i}\tau_{iy}S_{iz}$, which is obtained by projecting the Eq.~\eqref{SOC} in local S=1 and $\tau=1/2$ space. See SM~\cite{SM} for details. 
 
 The excitation spectrum calculated using LSOWT is shown in Fig.~\ref{fig4}(e). We fix the exchange parameters to match the spectrum to the RPA results in Fig~\ref{fig4}(d). The red and green color represents the $+1$ and $-1$ spin projections along $z$-direction $\langle S_{z}\rangle_{m{\bf q}}$, respectively, again depicting the chiral splitting in magnons. The spin projection is calulated using $\langle S_{z}\rangle_{m{\bf q}}=\sum_{i}\langle m {\bf q}| S_{iz}| m {\bf q}\rangle$ for band $m$. Notably, the spin polarization $\langle S_{z}\rangle_{m{\bf q}}$ exhibits the $d_{x^{2}-y^{2}}$-wave symmetry in momentum space even for non-zero SOC. This is attributed to the invariance of the ground state and the full $S$-$\tau$ Hamiltonian, including SOC, under the spin-space group operations $\{ {\mathcal{T}} | C_{4z} | T_{a}^{x/y}\}$, $\{ R_{x}(\pi) | m_{xy} | T_{a}^{x/y}\}$ and $\{ R_{x}(\pi) | m_{x{\bar{y}}} | T_{a}^{x/y}\}$~\cite{ZXiao01}. Here $T_{a}^{x/y}$ is the lattice translation and $R_{x}(\pi)$ is the spin-only rotation around $x$-axis; see also Table~\ref{table1} for the action of point group operations on pseudospins. It should be noted that magnetic space group treatment is not necessary here due to the nature of the weak SOC term in the $S$-$\tau$ model. The SOC induced spin gap $\Delta_{\rm gap}^{s}$ is also evident which we estimated as 
 \begin{equation}
 \Delta_{\rm gap}^{s}\approx 2\sqrt{(J^{S}-\frac{J_{1}^{S\tau}}{4}-2K)\lambda {\phi} }
 \end{equation}
 in the small tilt limit of ${\bm \tau}$ towards the $\pm y$ direction.
 
 The two weakly dispersive bands near $\omega=0.1$ are the pseudospin collective excitations, i.e., orbitons. The two orbiton modes correspond to the momenta ${\bf q}$ and ${\bf q}+M$ in the full Brillouin zone convention that we used throughout this work. Notably, the orbitons are dispersionless for $\lambda$=0 and any non-zero SOC induces canting in pseudospins, leading to dispersive orbitons from Ising exchange in $\tau_{z}$. We also confirmed that these orbitons do not carry net orbital polarization by calculating $\langle \tau_{z} \rangle_{m{\bf q}}=\sum_{i}\langle m {\bf q} | \tau_{iz}| m {\bf q}\rangle$, namely $\langle \tau_{z} \rangle_{m{\bf q}}=0$ for any ${\bf q}$.

{\subsection{Spin  conductivity}} 
 
 Having developed the LSOWT for the excited modes, we now estimate the magnon longitudinal and transverse spin conductivities in the presence of a thermal gradient, under the momentum-independent relaxation time approximation.  We use realistic values of the relaxation time of $\tau_{r}=10^{-10}$s for magnetic insulators~\cite{Mook01,Rezende01}. Details of the calculation are given in SM~\cite{SM}. The average temperature of $T\approx0.5T_{OO}$ is used. The critical temperatures of $T_{OO}\approx0.017k_{B}^{-1}$ and $T_{SO}\approx0.023k_{B}^{-1}$ are estimated by performing Monte-Carlo calculations on the $S$-$\tau$ model, for the same parameters used in Fig.~\ref{fig4}(e), see~\cite{SM} for details. The thermal gradient $\nabla T$ is chosen at an angle of $\eta$ with respect to the $x$-axis, see the inset of Fig.~\ref{fig4}(f). The longitudinal and transverse spin conductivities in dimensionless form ${\tilde{\sigma}}^{s}_{||(\perp)}=\frac{\hbar}{\tau_{r} k_{B} t_{z}}\sigma^{s}_{||(\perp)}$ are depicted in Fig.~\ref{fig4}(f). We find that the system shows the maximum longitudinal spin conductivity for thermal gradient along $\pm x$ or $\pm y$ directions, while the maximum transverse spin conductivity is attained if the thermal gradient is along the diagonal directions of the square lattice.  ${\tilde{\sigma}}^{s}_{||}$ and ${\tilde{\sigma}}^{s}_{\perp}$ are related to the spin conductivities in the $x$ and $y$ directions through 
 \begin{eqnarray}{\tilde{\sigma}}^{s}_{||}&=\tilde{\sigma}^{s}_{xx}\cos^{2}\eta + \tilde{\sigma}^{s}_{yy}\sin^{2}\eta,
 \nonumber
 \\
 {\tilde{\sigma}}^{s}_{\perp}&=(\tilde{\sigma}^{s}_{yy} -\tilde{\sigma}^{s}_{xx})\cos\eta\sin\eta.
 \end{eqnarray}
 The $d$-wave nature of spin polarization guaranties $\tilde{\sigma}^{s}_{yy}=- \tilde{\sigma}^{s}_{xx}$  suggesting ${\tilde{\sigma}}^{s}_{||}=2\tilde{\sigma}^{s}_{xx}\cos 2\eta$ and ${\tilde{\sigma}}^{s}_{\perp}=-2\tilde{\sigma}^{s}_{xx}\sin 2\eta$ which explains the results depicted in Fig.~\ref{fig4}(f). Fixing $\lambda=0.04t_{z}$, for the expected range of hopping $t_{z} \in \{ 0.1, 0.3\}$eV in real materials, we estimate that the maximum $\sigma_{||}^{s}$ or $\sigma_{\perp}^{s}$ can be of the order of (0.03-0.1)meV/K, while for a weaker SOC of $0.01t_{z}$ we expect larger spin conductivities (0.4-1.2)meV/K. The above suggests that in clean samples, the spin Seebeck and spin Nernst effects can be substantial and experimentally measurable~\cite{QCui01}.

~

\section{Magnon-orbiton hybridization}
The presence of orbiton modes close to magnons in energy raises an interesting possibility of hybridization between chiral magnons and orbitons in the altermagnetic state. The hybridization between the spin and pseudospin excitations for arbitrary momentum and energy can be quantified by the 
\begin{equation}
S^{+}\tau^{\alpha}({\bm q}, \omega)=-{\textrm{Im}}\langle gs| S_{{\bf q}}^{+} \frac{1}{\omega + \iota \eta -H + E_{0}} \tau_{-{\bf q}}^{\alpha}| gs\rangle,
\end{equation}
where $\alpha \in \{x,y,z\}$.
In the absence of SOC, the full spin rotational symmetry of the Hamiltonian prohibits the hybridization between spin excitations and orbitons of the collinear ground state. As discussed earlier, even in the presence of SOC represented as  $-\lambda\sum_{i}\tau_{iy}S_{iz}$, the spins remain collinear with the Neel vector along the $z$-direction. The $U(1)$ rotation symmetry in the $x$-$y$ plane is still present as $[S_{z},H]=0$. Under a spin rotation of $\pi$ about the $z$-axis, the ground state $|gs \rangle$ and Hamiltonian are invariant, and $S_{i}^{+} \rightarrow -S_{i}^{+}$. This implies  $S^{+}\tau^{\alpha}({\bm q}, \omega)=-S^{+}\tau^{\alpha}({\bm q}, \omega)$, showing that any magnon-orbiton hybrid mode is prohibited despite having finite SOC.

The $U(1)$ spin rotational symmetry discussed above can be explicitly broken by an external magnetic field in the $x$-$y$ plane. To that end, we include a small Zeeman field in $x$-direction by adding a term $-B_{x}\sum_{i}S_{ix}$ in the Hamiltonian. We discuss the effect of small $B_{x}$ and $\lambda$ within the Hartree-Fock approximation on the altermagnetic state. We find that, for any non-zero $B_{x}$, the spins in the ground state slightly cant at angle of $\theta$ in the $+x$-direction while maintaining the antiferro order in the $\pm z$-direction, leading to non-collinear spin state. Pseudospins meanwhile remain collinear in the $y$-$z$ plane with a small tilt towards $\pm y$ direction due to the finite SOC. The pictorial description of the state is shown in Fig.~\ref{fig5}(a,b).
We next performed the RPA calculations on this canted ground state.  In Fig.~\ref{fig5}(c) we show $S({\bf q}, \omega)$, for a small $B_{x}=0.005t_{z}$ while keeping the rest of the parameters the same as in Fig.~\ref{fig4}(d). Remarkably, we found that the two weakly dispersive orbiton modes near $\omega=0.1$ appear even for this purely spin dynamical structure factor, marking the presence of hybridization between the spin excitations and orbitons.

Furthermore, we confirmed the above predictions using the LSOWT on the modified $S$-$\tau$ model in the presence of finite $B_{x}$. The dispersion of collective modes is shown in Fig.~\ref{fig5}(e). Firstly, we describe the regions where orbitons cross the degenerate magnons, for example along the $\Gamma \leftrightarrow M$ line. The two degenerate magnons can be seen as the magnons corresponding to momentum ${\bf q}$ and ${\bf q}+M$ with both having zero $S_{z}$ projections, marking them as non-chiral. We found that the orbiton corresponding to momentum ${\bf q}$ $({\bf q}+M)$ only hybridizes with the magnon with the momentum ${\bf q}$ $({\bf q}+M)$, which keeps the hybrid magnon-orbiton mode non-chiral in nature, see Fig.~\ref{fig5}(d) for a zoomed-in view. This selective hybridization also keeps the spectrum gapless in the above regions as one of the magnon modes remains unhybridized. 
In contrast to the above, we found a qualitatively different type of hybridization between chiral magnons and orbitons. Notably, both the chiral magnons have equal contributions from  ${\bf q}$ and ${\bf q}+M$ momentum modes, hence the non-selective hybridization can occur for any orbiton branch that crosses a chiral magnon branch. In Fig.~\ref{fig5}(e) and (f), the color represents the orbital polarization, $\langle \tau_{z} \rangle_{m{\bf q}}$, of the Bloch states. Importantly, $\langle \tau_{z} \rangle_{m{\bf q}}$ is finite wherever orbitons cross near the chiral magnons. The enlarged view of a representative region where the orbitons and chiral magnons hybridize and open a gap is shown in Fig.~\ref{fig5}(f). In the presence of both non-zero $B_{x}$ and $\lambda$, the full $S$-$\tau$ model and the canted spin-pseudospin ground state that is depicted in Fig.~\ref{fig5}(a,b) are invariant under the spin-space group operations $\{ {\mathcal{T}}R_{z}(\pi) | C_{4z}|T_{a}^{x/y}\}$, $\{R_{x}(\pi) | m_{xy}|T_{a}^{x/y}\}$, and $\{ R_{x}(\pi) | m_{x{\bar{y}}}|T_{a}^{x/y}\}$, where  $R_{x(z)}(\pi)$ is the spin-only rotation around the $x(z)$-axis. Taking into account the above discussion and the table~\ref{table1}, within the LSOWT, it is expected that $\langle \tau_{z}\rangle_{m{\bf q}}=-\langle \tau_{z}\rangle_{{m}({ {\hat{\Theta}} \bf q})}$, where $ {\hat{\Theta}} \in \{ C_{4z}, m_{xy}, m_{x \bar{y}}\}$ suggesting the $d_{x^{2}-y^{2}}$ nature of orbital polarization, as also depicted in the inset of Fig.~\ref{fig5}(g) by plotting the $\langle \tau_{z}\rangle_{m{\bf q}}$ at an energy of 0.11.

Finally, we show that the hybrid mode of the orbiton and the chiral magnon gives rise to orbital-polarized conductivity under the thermal gradient. We calculate the orbital conductivity under the semiclassical approximation, as we did for spin conductivity. For details of the calculation, see SM~\cite{SM}. The dimensionless longitudinal conductivity in the $y$-direction $\tilde{\sigma}_{yy}^{\tau}=\frac{\hbar^{2}}{\tau_{r}k_{B}t_{z}}{\sigma}_{yy}^{\tau}$ as a function of $B_{x}$ is shown in Fig.~\ref{fig5}(g) and is consistent with the $\propto B_{x}^{2}$ behavior in the small $B_{x}$ regime. We also notice that $\tilde{\sigma}_{xx}^{\tau}=-\tilde{\sigma}_{yy}^{\tau}$, which directly follows from the $d$-wave momentum dependence of the orbital polarization. Due to the above, both transverse and longitudinal conductivities are also expected to exhibit a similar dependence on the thermal gradient direction as the magnon-based spin conductivity, see Fig.~\ref{fig4}(f). Importantly, it should be noted that the magnetic field of $\{0.005-0.01\}t_{z}$, for the realistic range of $t_{z} \in \{0.1,0.3\}$eV, and assuming the $g$-factor of 2, is equivalent to 4-24T which is an  accessible range of magnetic field for an experimental realization of the above described hybrid magnon-orbiton mode.

Findings discussed in this section have multiple intriguing implications. These include ($i$) possible observation of orbitons, which is generally challenging, directly in inelastic neutron scattering experiments, and ($ii$) possibility of exploiting the always present orbital-lattice coupling in generation of the hybrid magnon-orbiton mode via surface acoustic waves, by developing thin films of spontaneous altermagnets on a piezoelectric substrate~\cite{Taniguchi01,Gunnink01}. These could have applications in magnon-based spintronics. Finally, ($iii$) the hybrid chiral magnon-orbiton mode with finite $d$-wave type orbital polarization, as shown, is expected to give rise to orbital Seebeck and Nernst effects in the altermagnetic insulators, which can be advantageous for orbitronics.

\section{Summary and conclusions}

Emergence of spontaneous altermagnetism, whereby a system of interacting electrons spontaneously breaks both lattice and spin symmetries, requires violation of the well established Goodenough-Kanamori rules, which play a pivotal role in our understanding of quantum magnets. In this work, we introduced a microscopic mechanism that allows an interacting electron system with three active orbitals per site to circumvent the GK rules and support a staggered order in both spin and orbital degrees of freedom, thus forming a spontaneous altermagnet.  

In support of this new scenario we studied the stability criteria for spontaneous altermagnetism by employing an unbiased mean field technique on two-dimensional multi-orbital $t_{2g}^{2}$ systems and a DMRG method on the effecive large-$U$ spin-orbital model that we derived.  These considerations highlight three key ingredients necessary to stabilize the AFM+AFO state: ($i$) the tetragonally compressed environment, ($ii$) intermediate to strong Coulomb correlations, and ($iii$) robust intra-orbital hopping in the $xy$-orbital. Given that the above conditions are satisfied, we found the ALM state to be stable in a wide range of Hund's coupling at zero temperature. In addition, using classical Monte-Carlo simulations, we find that the Hund's coupling determines the sequence of the finite-temperature spin and orbital ordering phase transitions. We predict that spin and orbital-ordering critical temperatures depend oppositely on Hund’s coupling, that is, increasing Hund's coupling decreases the spin-ordering critical temperature while increasing the orbital-ordering temperature.

In terms of experimental signatures we make a prediction for chiral spin-split magnons in spontaneous altermagnets, based on the evaluation of the dynamical spin structure factor using the linearized time-dependent Hartree-Fock method. We confirm these results via the linear spin-orbital wave theory applied to our effective large-$U$ Hamiltonian, modified by adding higher-order terms constrained by symmetry. We furthermore predict substantial longitudinal and transverse spin conductivity rooted in the momentum-space spin polarization of chiral magnons. We also discuss the effect of weak atomic spin-orbit coupling on altermagnetism, spin excitations, and spin conductivity.

Finally, we predict that an in-plane magnetic field in the range of 4-24 T will hybridize magnons with orbitons thus realizing a novel hybrid chiral magnon-orbiton mode. Remarkably, this mode can be detected using conventional neutron  scattering and it also gives rise to measurable longitudinal and transverse orbital conductivity under a thermal gradient. 

Our results and predictions are directly relevant to quasi-two-dimensional materials with $t_{2g}^{2}$ active orbitals that satisfy the stability criteria discussed earlier. In particular, recent {\it ab-initio} calculations on layered perovskite materials, including CsVF$_{4}$~\cite{Lin01} and Sr$_{2}$CrO$_{4}$~\cite{Pandey01,Meier01}, suggest promising material platforms for the realization of the predicted AFO+AFM ground state. Other potential candidate materials include quasi-2D vanadates such as LnSrVO$_4$~\cite{Greedan01}, where epitaxial tensile strain may help achieve the required tetragonally compressed environment~\cite{Ricco01}. Notably,  $t_{2g}^{2}$ and $t_{2g}^{4}$ systems, with opposite tetragonal crystal fields, are related by particle-hole transformation within the $t_{2g}$ 
manifold. Hence, tetragonally elongated $t_{2g}^{4}$ systems are also expected to host spontaneous altermagnetism and the related phenomena discussed in the present work. Moreover, in real materials, the orbital degree of freedom is coupled to phonons via Jahn–Teller coupling. Consequently, a theoretical investigation of phonon effects constitutes an exciting direction for future work. Another promising avenue is the exploration of transport properties in electron- or hole-doped multi-orbital $t_{2g}^{2}$ systems; this could enable realization of spontaneous altermagnetic metals which hold promise for technological applications.

\section*{Acknowledgments} N. K. and M. F. were supported by NSERC, CIFAR and the Canada First Research Excellence Fund, Quantum Materials and Future Technologies Program. A.S.P. was supported by NSERC, CIFAR, and the Gordon and Betty Moore Foundation’s EPiQS Initiative through Grant No.\ GBMF11071 at the University of British Columbia.

~
~
\newpage
~
\newpage
\onecolumngrid{
\raggedbottom
\begin{center}
{\bf \uppercase{Supplementary Information}} for\\
\vspace{10pt}
{\bf \large Spontaneous altermagnetism in multi-orbital correlated electron systems}\\
\vspace{10pt}
by N. Kaushal, A. S. Patri, and M. Franz
\date{\today}
\end{center}
}

\twocolumngrid{
\setcounter{equation}{0}
\setcounter{figure}{0}
\setcounter{section}{0}
\setcounter{table}{0}
\setcounter{page}{1}
\makeatletter
\renewcommand{\theequation}{S\arabic{equation}}
\renewcommand{\thefigure}{S\arabic{figure}}

\maketitle

\section{Exact diagonalization}
\begin{figure*}[t]
\hspace*{-0.2cm}
\vspace*{0cm}
\includegraphics[width = 13.5cm]{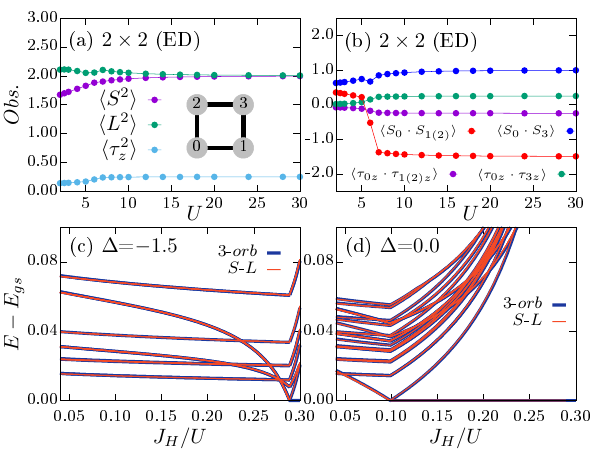}
\caption{Panels (a,b) depict the exact diagonalization results on a $2\times2$ cluster illustrated in the inset of panel (a). The averaged local spin moment $\langle S^{2}\rangle$, the averaged local orbital moment $\langle L^{2}\rangle$, and the averaged local pseudospin moment along $z$ direction $\langle \tau_{z}^{2}\rangle$ are shown as function of $U$ in panel (a). Panel (b) depicts the two-point spin-spin and pseudospin-pseudospin correlations for various bonds, again as a function of $U$. Panels (c) and (d) show the quantitative comparison between the low-energy spectrum of the 3-orbital Hubbard model and the derived $S-L$ model, for various $J_{H}/U$, for fixed $\Delta=-1.5$ and $\Delta=0.0$, respectively.}
\label{SMfig1}
\end{figure*}
In this section,  we discuss the exact diagonalization results on a $2\times 2$ three-orbital Hubbard model cluster, shown in the inset of Fig.~\ref{SMfig1}(a). We fix the parameters $J_{H}/U=0.2$, $\Delta=-1.5$, $t_{z}=1.0$, $t_{x}=0.7$, and tune $U$ from the weak to strong correlation regime. Fig.~\ref{SMfig1}(a) shows the evolution of the averaged local spin moment $\langle S^2 \rangle$, orbital moment $\langle L^2 \rangle$, and pseudospin moment along $z$ direction $\langle \tau_{z}^2 \rangle$ as a function of $U$. The averaged local observable is defined as $\langle O \rangle=(1/L_{x}L_{y})\sum_{i}\langle O_{i}\rangle$. In the large-$U$ limit, both $\langle S^2 \rangle$ and $\langle L^2 \rangle$ converges to approximately 2, indicating that the low-energy physics of the system can be described by $S=L=1$ local spaces. At the same time, $\tau_{z}^{2}$ converges to 0.25, consistent with local pseudospin $\tau=1/2$. Above observations demonstrate that the three-orbital Hubbard model in this regime can be described by the effective $S-L$ and $S-\tau$ models.

For completeness, we also present the spin-spin and pseudospin-pseudospin correlations in Fig.~\ref{SMfig1}(b). The sites ``0" and ``1(2)" exhibit antiferromagnetic spin and antiferro-orbital correlations for any $U>5$, whereas the correlations between sites ``0" and ``3" depict the ferromagnetic spin and ferro-orbital behavior for the same values of $U$.  The above AFM+AFO correlations are fully consistent with our results on larger systems obtained via Hartree-Fock and DMRG, as discussed in the main text.

\section{Perturbation theory and parameters of large-U limit effective models}
We used the Schrieffer-Wolf transformation to perform second-order perturbation theory in the limit $U>>t_{x/y/z}$ and to obtain the two-site low-energy effective Hamiltonian. Notably, the effective Hamiltonian for bonds along $y$ direction can be attained by acting $C_{4z}$ operation on the Hamiltonian attained for bonds along $x$ direction.
The elements of the effective Hamiltonian can be written as
\begin{equation}\label{SW}
H_{\textrm{eff}}^{\alpha \beta}= -\frac{1}{2} \sum_{n} H_{K}^{\alpha n} H_{K}^{n\beta}\left(\frac{1}{E_{n}-E_{\beta}} + \frac{1}{E_{n}-E_{\alpha}} \right),    
\end{equation}
where $H_{K}^{\alpha n}=\langle \alpha|H_{K}|n\rangle$ and $H_{K}^{n\beta}=\langle n|H_{K}|\beta\rangle$ are the elements of the multi-orbital kinetic energy term in the Hamiltonian. The $|\alpha (\beta)\rangle = |d^{2}_{\alpha(\beta)}\rangle_{0}|d^{2}_{\alpha(\beta)}\rangle_{1}  $ are the states in the ground state energy manifold of $H_{\textrm{int}}$ for two $t_{2g}^{2}$ sites, and $|n\rangle= |d_{n}^{1(3)}\rangle_{0} |d_{n}^{3(1)}\rangle_{1}$ represent the ground states for two sites with electronic configuration $t_{2g}^{1} t_{2g}^{3}$ or $t_{2g}^{3} t_{2g}^{1}$. 
For the convenience of the calculation, we exploit the fact that the total spin along $z$ direction $S_{z}$ is conserved, hence
\begin{equation}\label{SzProj}
H_{\textrm{eff}}=\sum_{m \in \{ 0, \pm1, \pm 2\}} H_{\textrm{eff},m},
\end{equation}
where $H_{\textrm{eff},m}=\hat{P}_{(S_{z}=m)} H_{\textrm{eff}} \hat{P}_{(S_{z}=m)}$ with  $\hat{P}_{(S_{z}=m)}=\sum_{\alpha}|m,\alpha\rangle \langle m, \alpha|$. The two-site state $|m,\alpha\rangle$ is a direct product state of ground states of $t_{2g}^{2}$ atoms , i.e., $|m' \alpha'\rangle_{0} | m'' \alpha'' \rangle_{1}$ where $\alpha',\alpha'' \in \{ x,y,z\}$, and $m',m'' \in \{ -1,0,1\}$ satisfying  $m'+m''=m$. In the above nomenclature, the state corresponding to site $i$, $|m\gamma\rangle_{i}=|m\rangle_{i}|\gamma\rangle_{i}$ represents a 2-electron state in $t_{2g}$ orbitals with $S_{z}=m$, $S=1$, and an empty orbital $\gamma$ with orbital indices $\{ x,y,z\} \equiv \{yz,xz,xy\}$. We use the equations~\ref{SW} and \ref{SzProj} to obtain elements of $H_{\textrm{eff}}$. To represent $H_{\textrm{eff}}$ in the operator form, we employ the following relations for each site,
\begin{fleqn}[0pt] 
\begin{multline}\label{OprAndStates}
|\gamma\rangle \langle\gamma| = 1 - L_{\gamma}^{2},
\\
|\gamma\rangle \langle\gamma'| = -L_{\gamma'}L_{\gamma}, {\textrm{for }} \gamma \ne \gamma', \hspace{0pt}
\\
|-1\rangle\langle -1| = 1 - \frac{S^{+}{S^{-}}}{2}, \hspace{0pt}
\\
|1\rangle\langle 1| = 1 - \frac{S^{-}{S^{+}}}{2}, \hspace{0pt}
\\
|0\rangle\langle 0| = 1 - {S_{z}^{2}}, \hspace{0pt}
\\
|1\rangle\langle 0| = \frac{S_{z}{S^{+}}}{\sqrt{2}}, \hspace{0pt}
\\
|-1\rangle\langle 0| = -\frac{S_{z}{S^{-}}}{\sqrt{2}}, {\textrm {and}} \hspace{0pt}
\\
|1\rangle\langle -1| = \frac{S^{+}{S^{+}}}{\sqrt{2}}. \hspace{141pt}
\end{multline}
\end{fleqn}
The final $S-L$ Hamiltonian obtained is discussed in detail in equations (3-9) of the main text. Below, we show the parameters of the $S-L$ model as functions of $J_{H}/U=\zeta$ and $U$,

\begin{fleqn}[0pt] 
\begin{multline*}
f_{1}^{SQ} = \frac{1}{U} \left( \frac{2\zeta}{1-3\zeta} + \frac{1+\zeta}{1+2\zeta} \right),
\hspace{0pt}
\\
f_{2}^{SQ} = -\frac{1}{U} \left( \frac{\zeta}{1-3\zeta} \right),
\hspace{0pt}
\\
f_{3}^{SQ} = \frac{1}{2U} \left( \frac{\zeta}{1+2\zeta} + \frac{1-\zeta}{1-3\zeta} \right),
\hspace{97pt}
\end{multline*}
\begin{multline}
f^{SL} = \frac{1}{2U} \left( \frac{1-\zeta}{1-3\zeta}-\frac{\zeta}{1+2\zeta} \right),
\hspace{0pt}
\\
f_{1}^{Q} = \frac{1}{U} \left( \frac{2(1-\zeta)}{1-3\zeta} - \frac{1+\zeta}{1+2\zeta} \right),
\hspace{0pt}
\\
f_{2}^{Q} = - \frac{1}{U} \left( \frac{1-\zeta}{1-3\zeta} \right),
\hspace{0pt}
\\
f_{3}^{Q} = \frac{1}{2U} \left( \frac{1+\zeta}{1-3\zeta} - \frac{\zeta}{1+2\zeta} \right), { \textrm{ and}}
\hspace{0pt}
\\
f^{L} = \frac{1}{2U} \left( \frac{1+\zeta}{(1-3\zeta)}+\frac{\zeta}{(1+2\zeta)} \right).
\hspace{67pt}
\end{multline}
\end{fleqn}
\newline

To confirm the validity of the derived $S-L$ model, we compare the low-energy spectrum of the two-site three-orbital Hubbard model with that of the corresponding two-site $S-L$ model, considering a bond along the $x$ direction. We fix $U=60$ to ensure the large-$U$ limit and tune the ratio $J_H/U$. The energy spectra of both models for $\Delta=-1.5$ and $\Delta=0.0$ are shown in Fig.~\ref{SMfig1}(c) and (d), respectively. The excellent quantitative agreement demonstrates that the $S-L$ model faithfully captures the low-energy physics of the three-orbital Hubbard model.

Moreover, for $\Delta<0$, in the limit of half-filled $xy$ orbital and the $xz/yz$ orbitals sharing the remaining electron, further simplification in the orbital part of $S$-$L$ model can be performed. For this purpose, we use $|z\rangle \langle z|={{1}}$, and $|\alpha\rangle \langle z|=0$ for $\alpha \in \{x,y\}$ to project the Hilbert space with half-filled $xy$ orbital. In addition to the above, we used the relations depicted in equations~\ref{OprAndStates}, and $\tau_{z}=(| y\rangle \langle y | - | x\rangle \langle x |)/2$. Above leads to $L_{z}^{2}={{1}}$, $Q^{\alpha z}=0$, and $\tau_{z}=(L_{x}^{2}-L_{y}^{2})/2$ which simplifies the $S-L$ model to $S-\tau$ model discussed in the main text. Below, we provide the exchange parameters of $H^{S\tau}$
\begin{fleqn}[0pt] 
\begin{multline}
J^{S} = \frac{t_{x/y}^{2}}{U}\left(-\frac{ \zeta }{2(1-3 \zeta )} + \frac{1+  \zeta  }{4(1+2  \zeta  )}\right)
+ \frac{t_{z}^{2}}{U}\frac{1+\zeta}{(1+2\zeta)},
\hspace{0pt}
\\
J_{1}^{S\tau} = \frac{t_{x/y}^{2}}{U} \left(\frac{2\zeta}{(1-3\zeta)} + \frac{1+\zeta}{(1+2\zeta)}\right),
\\
J_{2}^{S\tau} = \frac{t_{x/y}^{2}}{2U}  \frac{1+\zeta}{(1+2\zeta)}, {\textrm{ and}}
\hspace{0pt}
\\
J^{\tau} = \frac{t_{x/y}^{2}}{U} \left(\frac{2(1-\zeta)}{(1-3\zeta)} - \frac{1+\zeta}{(1+2\zeta)}\right).
\hspace{65pt}
\end{multline}
\end{fleqn}

Moreover, we also include the spin-orbit coupling in the $S-\tau$ model. The atomic spin-orbit coupling, see equation-14 of the main text, is projected to $S=1$ and $\tau=1/2$ space, which comprises only 6 local states corresponding to $S_{z} \in \{-1,0,1\} \times \tau_{z} \in \{-1/2,1/2\}$. Following the above, the effective SOC term can be written as
\begin{equation}
\hspace{0pt}
H_{SOC}= -\lambda\sum_{i}\tau_{iy}S_{iz}.
\end{equation}

\section{Details of linear spin-orbital wave theory}
\begin{figure*}[!t]
\hspace*{-0.2cm}
\vspace*{0cm}
\includegraphics[width = 13.5cm]{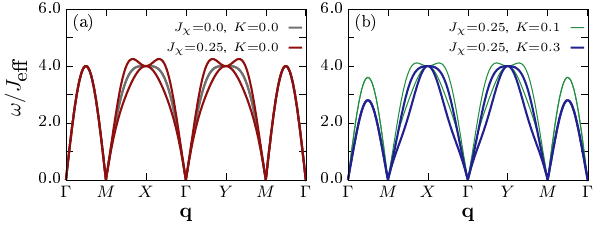}
\caption{Panel (a) depicts the magnon spectrum within the LSOWT for $(J_{\chi},K)=(0,0)$ and $(0.25,0)$, illustrating the chiral splitting induced by $J_{\chi}$. The effect of the ring exchange term $K$ on the magnon is shown in panel (b), by choosing the $(J_{\chi},K)=(0.25,0.1)$ and $(0.25,0.3)$. We fix $J_{\textrm{eff}}=J^{S}-\frac{J_{1}^{S\tau}}{4}$ to 1.0, for both the panels.}
\label{SMfig2}
\end{figure*}
In this section, we discuss the formalism employed to investigate the spin and orbital excitations. We used the full spin-pseudospin model described in the main paper, which includes the derived spin-pseudospin exchange terms in $H^{S\tau}$, the added chiral splitting $H_{\chi}$, the ring exchange $H_{R}$ terms, the atomic spin-orbit coupling, and also the in-plane magnetic field. We assume a two-sublattice spin-pseudospin canted ansatz state as pictorially depicted in Fig.6(a,b) of the main paper. The above direct-product state is parameterized by the canting angles $\theta$ and $\phi$, for the spin and pseudospin, respectively, about the AFM+AFO ordered state, and can be written as
\begin{multline*}
|\Psi(\theta,\phi)\rangle_{ans.} = (\Pi_{j\in A} |\theta \rangle_{j,S} |\phi \rangle_{j,\tau}) \otimes  (\Pi_{j\in B} |\theta \rangle_{j,S} |\phi \rangle_{j,\tau}).
\hspace{38pt}
\end{multline*}
The spin and pseudospin orientations are motivated by the results of our unrestricted Hartree-Fock calculations on a multi-orbital system, in the presence of the in-plane magnetic field and the spin-orbit coupling. The local spin vector, $(\langle S_{jx}\rangle, \langle S_{jy} \rangle, \langle S_{jz} \rangle) = {}_{j,S}\langle \theta | {\bf S}_{j} |\theta\rangle_{j,S}$, for the above state is considered to be $( \sin\theta, 0, \cos\theta )$, and $( \sin\theta, 0, -\cos\theta )$, for $j \in A$, and $B$, respectively. Similarly, the pseudospin vector  $\langle {\bm{\tau}}_{j} \rangle = {}_{j,\tau}\langle \phi | {\bm \tau}_{j} |\phi\rangle_{j,\tau}$ is assumed to be $1/2(0, \sin\phi, \cos\phi)$, and $1/2(0, -\sin\phi, -\cos\phi)$ for $j \in A$, and $B$, respectively. Using the above, we minimize the energy $_{ans.}\langle \Psi |H^{S\tau} + H_{\chi}+H_{R}+H_{SOC} + H_{B_{x}}| \Psi \rangle_{ans.}$ with respect to canting angles and attain the $\theta_{m}$ and $\phi_{m}$ representing the ground state.

We used the linear spin-orbital wave theory (LSOWT) to obtain the spin and orbital excitations on the above discussed canted state i.e. $| \Psi (\theta_{m}, \phi_{m}) \rangle$. To proceed further, we represent the local spin operators in terms of $SU(3)$ Schwinger bosons as 
\begin{equation}
\hspace{0pt}
S_{j\alpha}=-{\dot{\iota}} \sum_{\beta,\gamma}\epsilon_{\alpha \beta \gamma}a_{j\beta}^{\dagger}a_{j\gamma},
\end{equation}
where $\alpha,\beta,\gamma \in \{x,y,z\}$, with bosons satisfying the local constraint $\sum_{\alpha}a_{j\alpha}^{\dagger}a_{j\alpha}=M$ for any site $j$, where $M={\bf{S}}^{2}/2=1$. It should also be noted that the bosonic states, by acting ``$a$" operators on the vacuum, can be depicted as 
\begin{fleqn}[0pt] 
\begin{multline}
a_{jx}^{\dagger}|v\rangle_{j}^{a} \equiv ( {\dot{\iota}} |1\rangle_{j} -\iota |-1\rangle_{j})/\sqrt{2},
\\
a_{jy}^{\dagger}|v\rangle_{j}^{a} \equiv (|1\rangle_{j} + |-1\rangle_{j})/\sqrt{2}, {\textrm{ and}} 
\\
a_{jz}^{\dagger}|v\rangle_{j}^{a} \equiv - {\dot{\iota}} |0\rangle_{j},
\hspace{165pt}
\end{multline}
\end{fleqn}
where  $|v\rangle_{j}^{a}$ is the vacuum state of ``$a$" bosons for site $j$, and the states $\{ |\pm1\rangle_{j}, |0\rangle_{j} \}$ are the eigenstates of the operator $S_{jz}$. Using above, any local $S=1$ state $|{\mathbf d} \rangle_{j,S}$ can be uniquely represented using a complex vector ${\bf d}_{j}$ as $|{\mathbf d} \rangle_{j,S} = \sum_{\alpha} d_{j\alpha} a_{j\alpha}^{\dagger}|v\rangle_{j}^{a}$, where $d_{j\alpha}$ are the components of ${\bf d}_{j}$ vector. The vectors ${\bf d}=\{d_{x}, d_{y}, d_{z}\}$ representing the spin part, corresponding to the state $| \Psi(\theta_{m}, \phi_{m}) \rangle$, on the sublattice $A$ and $B$ are  ${\bf d}_{A}=1/\sqrt{2} ( \cos\theta_{m}, {\dot{\iota}}, -\sin\theta_{m} )$ and ${\bf d}_{B}=1/\sqrt{2}\{ -\cos\theta_{m}, {\dot{\iota}}, -\sin\theta_{m}\}$, respectively.

A linear combination of bare ``$a$" bosons is required to represent the spin-ordered state. Hence, we define the bosons $\tilde{a}_{j\alpha}$ in the site dependent rotated basis, using the rotation matrices $R_{jS}$,  by $[\tilde{a}_{j0}^{\dagger} \tilde{a}_{j1}^{\dagger} \tilde{a}_{j2}^{\dagger}]=[{a}_{jx}^{\dagger} {a}_{jy}^{\dagger} {a}_{jz}^{\dagger}]R_{jS}^{\dagger}$. The condensation of the bosons $\tilde{a}_{j0}$ corresponds to the spin ordering, with the first row of the rotation matrix $R_{jS}$ deciding the direction of the spin at site $j$. The other two rows, corresponding to $\tilde{a}_{j1(2)}$ bosonic excitations, of the matrix $R_{jS}$ can be arbitrarily chosen, while satisfying the condition $R_{jS}^{\dagger}R_{jS}=R_{jS}R_{jS}^{\dagger}=I$. The rotation matrices used for the sublattices A and B are
\begin{multline*}
R_{AS}=\frac{1}{\sqrt{2}}
\begin{bmatrix}
{\dot{\iota}} \cos\theta_{m} & 1 & - {\dot{\iota}} \sin\theta_{m} \\
-{\dot{\iota}} \cos\theta_{m} & 1 & {\dot{\iota}} \sin\theta_{m} \\
\sqrt{2}\sin\theta_{m} & 0 & \sqrt{2}\cos\theta_{m}
\end{bmatrix}
\hspace{100pt}
\end{multline*} and
\begin{multline*}
R_{BS}=\frac{1}{\sqrt{2}}
\begin{bmatrix}
-{\dot{\iota}} \cos\theta_{m} & 1 & -{\dot{\iota}} \sin\theta_{m} \\
{\dot{\iota}} \cos\theta_{m} & 1 & {\dot{\iota}} \sin\theta_{m} \\
-\sqrt{2}\sin\theta_{m} & 0 & \sqrt{2}\cos\theta_{m}
\end{bmatrix},
\hspace{100pt}
\end{multline*}
respectively.

Similarly, for the pseudospin space, we employ the $SU(2)$ Schwinger boson representation using the relation
\begin{equation}
\hspace{0pt}
{\bm{\tau}_{j}}=1/2\sum_{\alpha \beta}{\bm{ \sigma}}_{\alpha\beta}b_{j\alpha}^{\dagger}b_{j\beta},
\end{equation}
where $\alpha,\beta \in \{0,1\}$. Again, the number of bosons at each site satisfies the constraint $\sum_{\alpha}b_{j\alpha}^{\dagger}b_{j\alpha}=N$, where $N=4{\bm {\tau}}^{2}=1$. The local Hilbert space is spanned by the states $b_{j0}^{\dagger}|v\rangle_{j}^{b}=|1/2\rangle$, and  $b_{j1}^{\dagger}|v\rangle_{j}^{b}=|-1/2\rangle$. Again, we define bosons in the rotated basis by $[\tilde{b}_{j0}^{\dagger} \tilde{b}_{j1}^{\dagger}]=[{b}_{j0}^{\dagger} {b}_{j1}^{\dagger}] R_{j\tau}^{\dagger}$, where $\tilde{b}_{j0}$ and $\tilde{b}_{j1}$ represent the condensation channel and excitations, respectively. The rotation matrices used for the sites on sublattices A and B are
\begin{multline*}
R_{A\tau}=
\begin{bmatrix}
-\iota \cos\frac{\phi_{m}}{2} & -\sin\frac{\phi_{m}}{2} \\
\sin\frac{\phi_{m}}{2} & \iota \cos\frac{\phi_{m}}{2}
\end{bmatrix}
\hspace{100pt}
\end{multline*}
\\
and 
\\
\begin{multline*}
R_{B\tau}=
\begin{bmatrix}
-\iota \sin\frac{\phi_{m}}{2} & \cos\frac{\phi_{m}}{2} \\
-\cos\frac{\phi_{m}}{2} & \iota \sin\frac{\phi_{m}}{2}
\end{bmatrix}.
\hspace{100pt}
\end{multline*}
To perform the linear spin-orbital wave theory, firstly, we write the Hamiltonian in $\tilde{a}$ and $\tilde{b}$ bosons, exploiting the rotation matrices discussed above. Then, assuming the large M and N limit, we use ${\tilde{a}}_{j0}={\tilde{a}}_{j0}^{\dagger}\rightarrow\sqrt{M}(1 - \frac{{\tilde{a}}_{j1}^{\dagger}{\tilde{a}}_{j1} + {\tilde{a}}_{j2}^{\dagger}{\tilde{a}}_{j2} }{2M})$ and  ${\tilde{b}}_{j0}={\tilde{b}}_{j0}^{\dagger}\rightarrow\sqrt{N}(1 - \frac{{\tilde{b}}_{j1}^{\dagger}{\tilde{b}}_{j1} }{2N})$. Furthermore, to construct the LSOWT describing the small fluctuations around the ordered state, we only keep the quadratic terms. The obtained effective Hamiltonian in the momentum space is
\begin{fleqn}[0pt] 
\begin{multline}\label{LSWT}
H_{\textrm{LSOWT}}=\sum_{{\mathbf k}} \epsilon_{\mathbf{k}}^{a}\tilde{a}_{{\mathbf k}2}^{\dagger}\tilde{a}_{{\mathbf k}2} +\epsilon_{{\mathbf k}}^{b}\tilde{b}_{{\mathbf k}1}^{\dagger}\tilde{b}_{{\mathbf k}1} + \chi_{\mathbf k}\tilde{a}_{{\mathbf k}2}^{\dagger}\tilde{a}_{{\mathbf k}-M2}\\
+\Delta_{\bf k}^{a}(\tilde{a}_{{\bf k}2}\tilde{a}_{-{\bf k}2}+\textrm{h.c.})
+ \Delta_{\bf k}^{b}(\tilde{b}_{{\bf k}1}\tilde{b}_{-{\bf k}1}+\textrm{h.c.}) 
\\ + \Lambda^{ab}(\tilde{a}_{{\bf k}2}\tilde{b}_{-{\bf k}1} - \tilde{a}_{{\bf k}2}^{\dagger}\tilde{b}_{{\bf k}1} + \textrm{h.c.})
\hspace{70pt}
\end{multline}
\end{fleqn}
, where
\allowdisplaybreaks
\begin{fleqn}[0pt] 
\begin{align*}\label{H_LSOWT_PARAMS}
\epsilon_{\bf k}^{a}=-2\sin^{2}\theta_{m}(J^{S} - \frac{J_{1}^{S\tau}}{4}\cos^{2}\phi_{m})(\cos k_{x}+\cos k_{y}) 
\\
+4\cos2\theta_{m}(J^{S} - \frac{J_{1}^{S\tau}}{4}\cos^{2}\phi_{m}) + \frac{\lambda}{2}\sin\phi_{m}\cos\theta_{m}\\
+B_{x}\sin\theta_{m}+4K\cos2\theta_{m}\sin^{2}\theta_{m}(\cos k_{x} + \cos k_{y})
\\
-2K(\cos(k_{x}+k_{y})+\cos(k_{x}-k_{y}))-4K\cos4\theta_{m},
\\
\epsilon_{\bf k}^{b}=-\frac{\sin^{2}\phi_{m}}{2}(J_{\tau} - J_{1}^{S\tau}\cos2\theta_{m})(\cos k_{x}+\cos k_{y}) \hspace{10pt}\\
+ 2\cos^{2}\phi_{m}(J_{\tau} - J_{1}^{S\tau}\cos2\theta_{m}) + \lambda\sin\phi_{m}\cos\theta_{m},
\\
\chi_{\bf k}=J_{\chi}\cos\phi_{m}(\cos 2k_{x} - \cos 2k_{y}), 
\hspace{75pt}
\\
\Delta_{\bf k}^{a}=-\cos^{2}\theta_{m}(J^{S} - \frac{J_{1}^{S\tau}}{4}\cos^{2}\phi_{m})(\cos k_{x}+\cos k_{y})
\\
+2K\cos2\theta_{m}\cos^{2}\theta_{m}(\cos k_{x} + \cos k_{y}),
\hspace{36pt}
\\
\Delta_{\bf k}^{b}=-\frac{\sin^{2}\phi_{m}}{4}(J^{\tau} - {J_{1}^{S\tau}}\cos2\theta_{m})(\cos k_{x}+\cos k_{y}),
\\
\Lambda^{ab}=\frac{\lambda}{2\sqrt{2}} \cos\phi_{m}\sin\theta_{m}.
\hspace{114pt}
\end{align*}
\end{fleqn}
To derive the effective Hamiltonian described above, we set $N=M=1$. We also noticed that the excitations depicted the $a_{1}$ bosons are always gapped, even in the absence of spin-orbit coupling and external magnetic field, at an energy of about $8(J^{S}-\frac{J_{1}^{S\tau}}{4}-K)$ which is much higher than the energy of $a_{2}$ and $b_{1}$ excitations. Hence, for brevity in equation~\ref{LSWT}, we ignore the terms involving the $a_{1}$ bosons. We numerically perform the paraunitary diagonalization of the Bogoliubov Hamiltonian shown in eq.~\ref{LSWT}, following~\cite{Colpa01_SM}, to obtain,
\begin{equation}
H=\sum_{ \substack{\eta \in  \{0,1,2,3\} \\ {\bf k}   } }\omega_{\bf k}^{\eta}(\gamma_{{\bf k}\eta}^{\dagger}\gamma_{{\bf k}\eta} + \frac{1}{2})
\hspace{0pt}
\end{equation}
The eigenspectrum represented by $\omega_{\bf k}^{\eta}$ has already been addressed in detail in the main text. Here we discuss the limiting case of $B_{x}=0$ and $\lambda=0$. In this limit, $\phi_{m}=\theta_{m}=0$ leading to $\Lambda^{ab}=0$, hence we can safely ignore the orbiton-magnon hybridization part. The above simplification reduces the Hamiltonian into two separate parts, where the orbiton part exhibits flat dispersionless modes at an energy of  $2(J_{\tau}-J_{1}^{S\tau})$ and the magnon part that can be represented as below, 
\begin{multline}
H= \sum_{{\bf k}}(\Delta_{ {\bf k} }^{a}a_{ {\bf k} 2}^{\dagger}a_{- {\bf k} 2}^{\dagger} + {\textrm{h.c.}}) + \chi_{\bf k}a_{ {\bf k} 2}^{\dagger}a_{ {\bf k} -M2} + \epsilon_{\bf k}^{a}n_{ {\bf k} 2}
\hspace{0pt}
\end{multline}
where,
\begin{fleqn}[0pt] 
\begin{multline}
\hspace{0pt}
\Delta_{\bf k}^{a}=-(J_{\textrm{eff}} - 2K)(cos(k_{x})+cos(k_{y})), 
\\
\chi_{\bf k}=J_{\chi}(cos(2k_{x})-cos(2k_{y})),
\hspace{0pt}
\\
\epsilon_{\bf k}^{a}=4(J_{\textrm{eff}} - K) - 2K(cos(k_{x}+k_{y})+cos(k_{x}-k_{y})),\hspace{10pt}
\end{multline}
\end{fleqn}
and $J_{\textrm{eff}}=J^{S}-\frac{J_{1}^{S\tau}}{4}$.
We perform the para-unitary diagonalization of the above Hamiltonian to achieve two bosonic excitations with dispersions
\begin{equation}
\omega_{\bf k}^{\pm}=  \pm 
 \chi_{\bf k}+\sqrt{(\epsilon_{\bf k}^{a})^{2} - 4(\Delta_{\bf k}^{a})^{2}},
\end{equation}
where $\omega_{\bf k}^{+(-)}$ corresponds to positive (negative) chiral magnons. We depict the magnon spectrum in Fig.~\ref{SMfig2} for various values of $(J_{\chi},K)$, fixing $J_{\textrm{eff}}=1.0$. In Fig.~\ref{SMfig2}(a), we illustrate the chiral splitting in magnons induced by the non-zero $J_{\chi}$, however, it also shifts the maxima away from the ${\mathbf{q}}=X/Y$ points. We find that adding the large enough ring exchange term, which is also expected to emerge from fourth order perturbation theory, can retain the maxima at ${\mathbf{q}}=X/Y$ points, see Fig.~\ref{SMfig2}(b). Moreover, the ring exchange term softens the magnon energy along the $\Gamma-M$ paths, and eventually makes it resemble our RPA results.

\section{Spin and Orbiton conductivity calculations}
The collective modes calculated by the LSOWT can be used to estimate the spin and orbital conductivities. Using the semi-classical Boltzmann equation and employing the steady state and relaxation time approximations, the spin conductivity driven by spin-polarized magnons can be written as~\cite{QCui01_SM},
\begin{equation}\label{Spincond}
\sigma_{\alpha\alpha}^{s}=-\frac{\tau_{r} \hbar}{A}\sum_{ {\mathbf k} m}\langle S_{z}\rangle_{m {\mathbf k}}\frac{\partial n^{0}_{m{\bf k}}}{\partial T} v_{m {\bf k}}^{\alpha} v_{m {\bf k}}^{\alpha},
\end{equation}
where the $v_{m {\bf k}}^{\alpha}=\frac{1}{\hbar}\frac{\partial \epsilon_{m \bf k}}{\partial k_{\alpha}}$, $\tau_{r}$ is the momentum independent relaxation time, and $n^{0}_{m{\bf k}}=\frac{1}{e^{\beta\epsilon_{m {\bf k}}}-1}$. Based on the above, we can write $\sigma_{\alpha\alpha}^{s}=\frac{\tau_{r}k_{B}t_{z}}{\hbar} {\tilde{\sigma}}_{\alpha\alpha}^{s}$ , where ${\tilde{\sigma}}_{\alpha\alpha}^{s}$ is dimensionless and is represented in the following manner,
\begin{equation}
{\tilde{\sigma}}_{\alpha\alpha}^{s} = - \frac{1}{ A}\sum_{{\bf k} m}\langle S_{z}\rangle_{m {\mathbf k}} \frac{\epsilon_{m{\bf k}}}{t_{z}} \frac{e^{\beta\epsilon_{m{\bf k}}}} { (e^{\beta\epsilon_{m{\bf k}}} -1)^{2} } \left(\frac{\partial (\beta \epsilon_{m {\bf k} })}{\partial k_{\alpha}}\right)^{2}
\end{equation}
where, $\beta=(k_{B}T)^{-1}$, and $T$ is the average temperature of the sample.  Note that we keep $t_{z}$ as a unit of energy. The above representation helps to easily estimate the spin conductivity for various $t_{z}$'s while keeping the rest of the model parameters, including the $k_{B}T$, fixed with respect to $t_{z}$. We use a reasonable value of relaxation time of magnons in clean magnetic insulators that is $\tau_{r}=10^{-10}$s. 

Similarly, we can write the orbital conductivity driven by the orbital polarized collective modes. To this end, we write,
\begin{equation}
\sigma_{\alpha\alpha}^{\tau}=-\frac{\tau_{r}}{A}\sum_{ {\mathbf k} m}\langle \tau_{z}\rangle_{m {\mathbf k}}\frac{\partial n^{0}_{m{\bf k}}}{\partial T} v_{m {\bf k}}^{\alpha} v_{m {\bf k}}^{\alpha}.
\hspace{0pt}
\end{equation}
We emphasize that we use the orbital polarization operator $\tau_{z}$, not the orbital angular momentum, hence $\hbar$ does not appear as in eq.~\ref{Spincond}. Nonetheless, we use the dimensionless form of orbital conductivity ${\tilde{\sigma}}_{\alpha\alpha}^{\tau}$, estimated using $\sigma_{\alpha\alpha}^{\tau}=\frac{\tau_{r}k_{B}t_{z}}{ {\hbar}^{2} } {\tilde{\sigma}}_{\alpha\alpha}^{\tau}$,  to present our results in the main text

\section{Linear Response in time-dependent Hartree Fock}
As discussed in the main text we used the unrestricted Hartree-Fock technique to solve the three-orbital Hubbard model. We assumed that the single particle density matrix, $\rho^{i}_{\alpha \alpha'}=\langle c_{i\alpha}^{\dagger}c_{i{\alpha}'}\rangle$, respects the reduced translational symmetry by satisfying  $\rho_{\alpha \alpha'}({\mathbf r}_{i} + m_{1}{\mathbf a}_{m1} + m_{2}{\mathbf a}_{m2})= \rho_{\alpha \alpha'}({\mathbf r}_{i})$, where ${\bf r}_i$ is the position of atom $i$, $(\alpha,\alpha')$ are the combined spin and orbital degree of freedoms, and ${\mathbf{a}_{1(2)m}}$  and $m_{1(2)}$ are the magnetic Bravais lattice vectors and arbitrary integers, respectively. To attain the lowest energy self-consistent solution, Hartree-Fock calculation is performed for various magnetic unit cells sizes upto the full system size starting from multiple random initial guesses. Considering a given translational symmetry in converged order parameters, the interaction part of the multi-orbital Hubbard model within Hartree-Fock approximation can be written as $H_{HF}^{Int}=\sum_{\bf k} H_{HF}^{Int}( {\bf k} )$
, where 
\begin{equation}
H_{HF}^{Int}({\bf k})=\sum_{\substack{  { n \alpha}{\alpha{'}} }} V^{n}_{ \alpha \alpha{'} }[\rho^{n}] c_{{\bf k} n \alpha }^{\dagger}c_{{\bf k} n \alpha{'}  }.
\hspace{0pt}
\end{equation}
The ${\bf k}$, $n$, and $\alpha(\alpha{'})$ are the momentum point in the magnetic Brillouin zone (MBZ), the atom position inside the magnetic unit cell (MUC), and the combined orbital and spin indices, respectively. The interaction matrix elements $V^{n}_{\alpha \alpha{'} }[\rho^{n}]$ depend on the single particle density matrix, but not on momentum because of local nature of the interaction.
\begin{figure*}[!t]
\hspace*{-0.2cm}
\vspace*{0cm}
\includegraphics[width = 13.5cm]{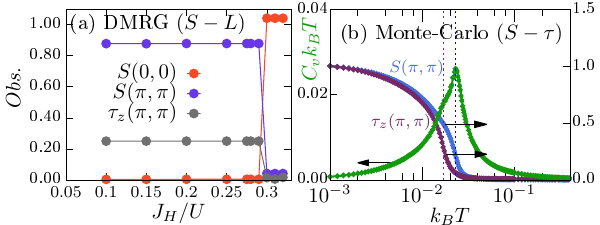}
\caption{Panel (a) depicts static structure factor values calculated using DMRG on $8\times 4$ $S-L$ model for fixed $t_{x}=0.6t_{z}$, $U=20t_{z}$, and $\Delta=-1.5t_{z}$. Results of $CP_{2}\times O_{3}$ Monte Carlo simulations on $S-\tau$ model, for $12\times12$ lattice, are depicted in panel (b).}
\label{SMfig3}
\end{figure*}

The density-matrix–to–density-matrix linear response is evaluated under the assumption of a weak time-periodic external potential linearly coupled to the density matrix~\cite{Giuliani01_SM}, leading to the following form:
\begin{multline}\label{ChiHF}
\chi^{HF}_{[{\hat{\rho}_{\alpha\alpha'}^{-{\bf k}n},\hat{\rho}_{\tilde{\alpha}{\tilde{\alpha}}'}^{{\bf k}n' }}]}({\bf k},\omega)=\sum_{ \substack{m \mu \mu'} } [{{1}} - \chi^{0}({\bf k},\omega)\cdot I]^{-1}_{n\alpha\alpha',m\nu\nu'} \times
\\
\chi^{0}_{m\nu\nu',n'\tilde{\alpha}\tilde{\alpha}'}({\bf k},\omega),
\end{multline}
where $\hat{\rho }_{\alpha\alpha'}^{{\bf k}n}=\sum_{l}e^{\dot{\iota}{\bf k} \cdot {\bf r}_{l}}c^{\dagger}_{ln\alpha}c^{}_{ln\alpha'}$ and $l$ is the MUC index. The indices $(n,n',m)$ and $(\alpha,\alpha',\tilde{\alpha},\tilde{\alpha}',\nu,\nu')$ correspond to atom positions in MUC and combined spin and orbital degree of freedoms, respectively. The $\chi^{0}({\bf k},\omega)$ is the bare response function calculated using Hartree-Fock eigenspectrum as below,
\begin{multline}\label{ChiBare}
\chi^{0}_{m\nu\nu',n'\tilde{\alpha}\tilde{\alpha}'}({\bf k},\omega) = \frac{1}{N_{muc}}\sum_{\lambda \lambda' {\bf p}} \left( \frac{ f_{{\bf p}\lambda} - f_{ {\bf p}+{\bf k} \lambda' } }{  \omega +\epsilon_{ {\bf p}\lambda}  - \epsilon_{ {\bf p}+{\bf k} \lambda'} + \dot{\iota}\eta} \right)
\\
\times (\psi_{ {\bf p} m \nu }^{\lambda})^{*} \psi_{ {\bf p} + {\bf k} m \nu' }^{\lambda'} (\psi_{ {\bf p} + {\bf k} n' \tilde{\alpha}}^{\lambda'})^{*} \psi_{ {\bf p} n' \tilde{\alpha}'}^{\lambda},
\end{multline}
where $N_{muc}$ is the number of magnetic unit cells, and $\epsilon_{ {\bf p} \lambda }$ and $\psi_{ {\bf p} m\nu}^{\lambda}$ are the $\lambda$th eigenvalue and eigenfunction, respectively, of $H_{HF}({\bf p}) = H_{KE}({\bf p}) + H^{Int}_{HF}({\bf p})$, and $f_{ {\bf p}\lambda } = \frac{1}{e^{(\epsilon_{ {\bf p} \lambda }-\mu) } +1 }$ represents the Fermi function. The interaction kernel ($I$), used in eq.~\ref{ChiHF}, is defined as 
\begin{equation}
I_{{n \alpha \alpha' },{ m \nu \nu' }}=\frac{\delta V^{n}_{\alpha \alpha'}[\rho^{n}]}{\delta \rho^{m}_{\nu \nu'}}\delta_{nm}
\hspace{0pt}
\end{equation}
Now onwards, we keep the notation $I_{{n \alpha \alpha' },{ n \nu \nu' }}=I_{{ (\alpha,\sigma) (\alpha',\sigma') },{ (\nu,s) (\nu',s') }}$ to explicitly include the spin indices using $(\sigma,\sigma',s,s')$ and remove the index $n$ as the interaction kernel is invariant under translation. We calculate the interaction kernel elements using the multiorbital local interaction, which are depicted below term by term,
\begin{fleqn}[0pt]
\begin{multline}
{\textrm{Intra-orbital repulsion term:}}
\\
I_{{ (\alpha,\sigma) (\alpha,\sigma) },{ (\alpha,\bar{\sigma}) (\alpha, \bar{\sigma}) }}=U
\\
I_{{ (\alpha,\sigma) (\alpha,\bar{\sigma}) },{ (\alpha,\bar{\sigma}) (\alpha, {\sigma}) }}=-U
\\
\\
{\textrm{Inter-orbital repulsion term:}}
\\
I_{{ (\alpha,\sigma) (\alpha,\sigma) },{ (\beta,{\sigma}') (\beta, {\sigma}') }}=U' - \frac{J_{H}}{2}
\\
I_{{ (\beta,\sigma) (\alpha,\sigma) },{ (\alpha,{\sigma}) (\beta, {\sigma}) }}=-(U' - \frac{J_{H}}{2})
\\
I_{{ (\beta,\sigma) (\alpha,\bar{\sigma}) },{ (\alpha,{\bar{\sigma}}) (\beta, {\sigma}) }}=-(U' - \frac{J_{H}}{2})
\\
\\
{\textrm{Hund's coupling term:}}
\\
I_{{ (\beta,{\bar{\sigma}}) (\beta,{\sigma}) },{ (\alpha,{{\sigma}}) (\alpha, {\bar{\sigma} }) }}=-J_{H}
\\
I_{{ (\beta,{{\sigma}}) (\beta,{\sigma}) },{ (\alpha,{{\sigma}}) (\alpha, {{\sigma} }) }}=-J_{H}/2
\\
I_{{ (\beta,{{\sigma}}) (\beta,{\sigma}) },{ (\alpha,{\bar{\sigma}}) (\alpha, {\bar{\sigma} }) }}=J_{H}/2
\\
I_{{ (\beta,{{\sigma}}) (\alpha,{\sigma}) },{ (\alpha,{\bar{\sigma}}) (\beta, {\bar{\sigma} }) }}=J_{H}
\\
I_{{ (\beta,{{\sigma}}) (\alpha,{\sigma}) },{ (\alpha,{{\sigma}}) (\beta, {{\sigma} }) }}=J_{H}/2
\\
I_{{ (\beta,{{\sigma}}) (\alpha, {\bar{\sigma}}) },{ (\alpha,{\bar{\sigma}}) (\beta, {{\sigma} }) }}=-J_{H}/2
\\
\\
\\
{\textrm{Pair hopping term:}}
\\
I_{{ (\alpha,{\bar{\sigma}}) (\beta,{\sigma}) },{ (\alpha,{{\sigma}}) (\beta, {\bar{\sigma} }) }}=-J_{H}
\\
I_{{ (\alpha,{{\sigma}}) (\beta,{\sigma}) },{ (\alpha,{\bar{\sigma}}) (\beta, {\bar{\sigma} }) }}=J_{H}
\\
\end{multline}
\end{fleqn}
Using the interaction Kernel described above, eq.~\ref{ChiBare}, and eq.~\ref{ChiHF}, we calculate $\chi^{HF}({\bf k},\omega)$. 

The response function between any two operators $A$ and $B$, which can be written as a linear combination of the single particle density operator as
\begin{equation}
O_{{\bf k}n}=\sum_{\alpha \beta} O^{\alpha}_{\beta} \hat{\rho}_{\alpha \beta}^{ {\bf k}n},
\end{equation}
can be described using
\begin{equation}
\chi^{HF}_{ [A_{-{\bf k}n},B_{{\bf k}n'}] }= \sum_{\alpha\beta\alpha'\beta'}A_{\beta}^{\alpha}B_{\beta'}^{\alpha'}\chi^{HF}_{[ \hat{\rho}_{\alpha \beta}^{ -{\bf k} n }, \hat{\rho}_{\alpha' \beta'}^{ {\bf k} n' } ]}.
\end{equation}
At last, the dynamical correlation function between A and B, in the original Brillouin zone, is evaluated using the following sum over the $N_{a}$ atoms inside the MUC
\begin{equation}\label{DCF}
AB( {\bf k}, \omega ) = {\textrm{Im}} \frac{1}{N_{a}}\sum_{n m}e^{\dot{\iota} {\bf k}\cdot( {\bf r}_{n} - {\bf r}_{m})}\chi^{HF}_{[A_{-{\bf k}n},B_{{\bf k}n'}]}( {\bf k}, \omega ).
\end{equation}
For the main text, the total spin dynamical structure factor $S({\bf k},\omega)$ and the measure of chiral splitting $M_{\chi}( {\bf k},\omega)$ are calculated employing eq.~\ref{DCF} using the corresponding $A$ and $B$ operators.

\section{low $t_{x}$ DMRG results}
In this section, we discuss the DMRG results on the $8\times 4$ $S-L$ model for $t_{x}=0.6t_{z}$,  $U=20t_{z}$ and $\Delta=-1.5t_{z}$. The evolution of static structure factors $S({\bf q})$, for ${\bf q}=(0,0)$ and $(\pi,\pi)$, and $\tau_{z}(\pi,\pi)$ as a function of $J_{H}/U$ is shown in Fig.~\ref{SMfig3}(a). The structure factors are calculated using the $6\times4$ lattice embedded inside the $8\times4$ lattice to minimize the open boundary effects of cylindrical geometry.
We notice that AFM+AFO state, depicted by robust $S(\pi,\pi)$ and $\tau_{z}(\pi,\pi)$, is stable in large range of $J_{H}/U$ and until $J_{H}/U \approx 0.28$. For larger $J_{H}/U$, we notice the FM+OD state illustrated by a large $S(0,0)$ and negligible $\tau_{z}({\bf q})$. In particular, compared to the larger $t_{x}=t_{z}$ discussed in the main text, $t_{x}=0.6t_{z}$, evidently, shows a similar behavior of transiting from AFM to FM state as $J_{H}$ is increased. However, (1) the intermediate state of FM+AFO is absent in $t_{x}/t_{z}=0.6$, and (2) the AFO+AFM state is stable relatively until larger values of $J_{H}$ for smaller $t_{x}/t_{z}$, which is consistent with the discussion on $S-\tau$ model provided in the main text.

\section{Details of Classical Monte-Carlo}
In this section, we discuss the details of the classical Monte-Carlo calculations performed on both the $S-L$ model and the $S-\tau$ model. In the main text, we already discussed the results corresponding to the $S-L$ model for various Hund's coupling strengths. Here, we provide the details of the ${\mathbb{CP}}^{2} \times {\mathbb{CP}}^{2}$ space representation, which is employed to capture the local $S=1$ and $L=1$ spaces faithfully.

Any local $S(L)=1$ state, for site $j$, can be represented by using the basis $\{|x\rangle, |y\rangle, |z\rangle\}$, where
\begin{multline}
|x\rangle=\frac{\dot{\iota}}{\sqrt{2}}(|1\rangle - |-1\rangle),
|y\rangle=\frac{1}{\sqrt{2}}(|1\rangle + |-1\rangle)
,
|z\rangle= -\dot{\iota}|0\rangle,
\end{multline}
as 
\begin{equation}\label{localdState}
|{\bf d}_{j}^{S(L)}\rangle_{jS(L)}=\sum_{\alpha \in \{x,y,z\}}d_{j\alpha}^{S(L)} |\alpha \rangle_{jS(L)}.
\end{equation}
The coefficients $d_{j\alpha}^{S(L)} \in {\mathbb{C}}$ uniquely specify the state at site $j$. Under a semiclassical approximation, we assume that the full state on the lattice can be written as the following direct product state
\begin{equation}
|\Psi\rangle = \Pi_{j}|{\bf d}_{j}^{S}\rangle_{jS}|{\bf d}_{j}^{L}\rangle_{jL}.
\end{equation}
To proceed further, following~\cite{Remund01_SM}, we use the following mapping of the vector ${\mathbf{d}}_{j}^{S(L)}=(d_{jx}^{S(L)}, d_{jy}^{S(L)}, d_{jz}^{S(L)})$ to $\mathbb{CP}^{2}$ space spanned by 4 independent variables as shown below,
\begin{fleqn}[0pt]
\begin{multline}
 \operatorname{Re}(d_{jx}^{S(L)})= \theta_{2S(L)}^{1/4}\theta_{1S(L)}^{1/2} \sin{\phi_{1S(L)}},
 \\
 \operatorname{Im}(d_{jx}^{S(L)})= \theta_{2S(L)}^{1/4}\theta_{1S(L)}^{1/2} \cos{\phi_{1S(L)}},
 \\
 \operatorname{Re}(d_{jy}^{S(L)})= \theta_{2S(L)}^{1/4}  \sqrt{1-\theta_{1S(L)}}  \sin{\phi_{2S(L)}},
 \\
 \operatorname{Im}(d_{jy}^{S(L)})= \theta_{2S(L)}^{1/4}  \sqrt{1-\theta_{1S(L)}}  \cos{\phi_{2S(L)}},
 \\
 \operatorname{Re}(d_{jz}^{S(L)})= \sqrt{1-\theta_{2S(L)^{1/2}}},
 \\
 \operatorname{Im}(d_{jz}^{S(L)})=0.
 \hspace{147pt}
\end{multline}
\end{fleqn}
In the above construction, the constraints $0 \leq \theta_{j1}^{S(L)}, \theta_{j2}^{S(L)} \leq 1$ and $0 \leq \phi_{j1}^{S(L)}, \phi_{j2}^{S(L)}  \leq 2\pi$ are satisfied to ensure that $|{\mathbf{d}}_{j}^{S(L)}|^{2}=1$ faithfully representing $S=L=1$ spaces. 
Importantly, the $\operatorname{Im}(d_{jz}^{S(L)})=0$ is fixed by choice, which is equivalent to multiplying each local state in eq.~\ref{localdState} with just a phase factor of $(d_{jz}^{S(L)})^{*}/|d_{jz}^{S(L)}|$.

We perform the classical Monte-Carlo using the Metropolis algorithm to accept/reject the random updates on the variables discussed above. The thermal annealing was performed with 20000 Monte-Carlo sweeps on the full lattice for each temperature point. Moreover, to calculate any observables, the ensemble averaging over 10000 microstates was performed after attaining thermalization.
For the specific heat $C_{v}$, we used
\begin{equation}
C_{v}=\frac{\langle E^{2}\rangle - \langle E \rangle^{2}}{k_{B}T^{2}},
\end{equation}
where $\langle \rangle$ represent the ensemble averaging.

As discussed in the main text, we also performed the Monte Carlo on $S-\tau$ model. To this end, we retain the $\mathbb{CP}^{2}$ representation for $S=1$ space; however, $\tau=1/2$ states can be represented by $O_{3}$ classical vectors as $\{\tau_{x}, \tau_y, \tau_{z} \}= \{ \sin\theta_{\tau} \cos\phi_{\tau}, \sin\theta_{\tau} \sin\phi_{\tau}, \cos\theta_{\tau} \}$ under the semiclassical approximation. We fixed the parameters $J^{S}$=0.052, $J_{1}^{S\tau}$=0.054, $J^{\tau}$=0.108, $J_{\chi}$=0.005, $K$=0.0056, and  $\lambda$=0.04, and perfomed the Monte Carlo within $\mathbb{CP}^{2} \times O_{3}$ representation.  The spin and pseudospin static structure factors at momentum $(\pi,\pi)$, and the specific heat as a function of temperature are shown in Fig.~\ref{SMfig3}(b). The two vertical dashed lines at temperatures $0.017k_{B}$ and $0.023k_{B}$ correspond to the critical temperatures of antiferro-orbital ordering and antiferro-spin ordering, respectively. Transition temperatures are identified from peaks in the specific heat and evolution of static structure factors.


}

\begin{thebibliography}{10}
\bibitem{Tokura01} Y. Tokura and N. Nagaosa, {\href{https://www.science.org/doi/10.1126/science.288.5465.462}{Science {\bf 288},462-468 (2000)}}.
\bibitem{Si01} Q. Si, R. Yu, and E. Abrahams, {\href{https://doi.org/10.1038/natrevmats.2016.17}{Nat. Rev. Mater. {\bf 1}, 16017 (2016)}}.
\bibitem{Daghofer01} M. Daghofer, A. Nicholson, A. Moreo, and E. Dagotto, {\href{https://journals.aps.org/prb/abstract/10.1103/PhysRevB.81.014511}{Phys. Rev. B {\bf 81}, 014511 (2010)}}.
\bibitem{DagottoCMR} E. Dagotto, T. Hotta, and A. Moreo, {\href{https://www.sciencedirect.com/science/article/abs/pii/S0370157300001216}{Phys. Rep. {\bf 344}, 1 (2001)}}.
\bibitem{Anisimov01} V. Anisimov, I. Nekrasov, and D. Kondakov, {\href{https://doi.org/10.1140/epjb/e20020021}{Eur. Phys. J. B {\bf 25}, 191–201 (2002)}}.

\bibitem{Khaliullin01} G. Khaliullin, {\href{https://journals.aps.org/prl/abstract/10.1103/PhysRevLett.111.197201}{Phys. Rev. Lett. {\bf 111}, 197201 (2013)}}.
\bibitem{Kaushal01} N. Kaushal, J. Herbrych, A. Nocera, G. Alvarez, A. Moreo, F. A. Reboredo, and E. Dagotto, {\href{https://journals.aps.org/prb/abstract/10.1103/PhysRevB.96.155111}{Phys. Rev. B {\bf 96}, 155111 (2017)}}.
\bibitem{Khomskii01} D.I. Khomskii and K.I. Kugel, {\href{https://www.sciencedirect.com/science/article/pii/0038109873903621?via%3Dihub}{l, Solid State Commun. {\bf 13}, 763 (1973)}}.

\bibitem{KhomskiiBook} D. I. Khomskii,  {\href{https://doi.org/10.1017/CBO9781139096782}{Transition Metal Compounds, Cambridge Univ Press, Cambridge, UK (2014)}}.

\bibitem{Pavarini01} E. Pavarini and Erik Koch,  {\href{https://journals.aps.org/prl/abstract/10.1103/PhysRevLett.104.086402}{Phys. Rev. Lett. {\bf 104}, 086402 (2010)}}.
\bibitem{Liechtenstein01} A. I. Liechtenstein,  V. I. Anisimov, J. Zaanen, {\href{https://journals.aps.org/prb/abstract/10.1103/PhysRevB.52.R5467}{Phys. Rev. B {\bf 52}, R5467(R) (1995)}}.

\bibitem{Itol01} Y. Itol, and J. Akimitsul, {\href{https://journals.jps.jp/doi/10.1143/JPSJ.40.1333?mobileUi=0}{J. Phys. Soc. Jpn. 40, pp. 1333-1338 (1976)}}.

\bibitem{Hayami01} S. Hayami, Y. Yanagi, and H. Kusunose, {\href{https://journals.jps.jp/doi/10.7566/JPSJ.88.123702}{J. Phys. Soc. Jpn. {\bf 88}, 123702 (2019)}}.
\bibitem{Smejkal01} L. {\v{S}}mejkal, R. Gonz\'alez-Hern\'andez, T. Jungwirth, and
J. Sinova, {\href{https://www.science.org/doi/10.1126/sciadv.aaz8809}{Sci. Adv. {\bf 6}, eaaz8809 (2020)}}.
\bibitem{Yuan01} L.-D. Yuan, Z. Wang, J.-W. Luo, E. I. Rashba, and A. Zunger, {\href{https://journals.aps.org/prb/abstract/10.1103/PhysRevB.102.014422}{Phys. Rev. B {\bf 102}, 014422 (2020)}}.
\bibitem{Mazin01} I. I. Mazin, K. Koepernik, M. D. Johannes, R. Gonz\'alez-Hern\'andez, and L. {\v{S}}mejkal , {\href{https://www.pnas.org/doi/full/10.1073/pnas.2108924118}{Proc. Natl. Acad. Sci. U.S.A. {\bf 118}, e2108924118 (2021)}}.


\bibitem{Leeb01}V. Leeb, A. Mook, L. {\v{S}}mejkal, and J. Knolle, {\href{https://journals.aps.org/prl/abstract/10.1103/PhysRevLett.132.236701}{Phys. Rev. Lett. {\bf 132}, 236701 (2024)}}.

\bibitem{YLi01} Y.-X. Li and C.-C. Liu, {\href{https://journals.aps.org/prb/abstract/10.1103/PhysRevB.108.205410}{Phys. Rev. B {\bf 108}, 205410 (2023)}}.
\bibitem{Zhu01} D. Zhu, Z.-Y. Zhuang, Z. Wu, and Z. Yan, {\href{https://journals.aps.org/prb/abstract/10.1103/PhysRevB.108.184505}{Phys. Rev. B {\bf 108}, 184505 (2023)}}.

\bibitem{Monkman01} K. Monkman, J. Weng, N. Heinsdorf, A. Nocera, and M. Franz, {\href{http://arxiv.org/abs/2507.22139}{arXiv:2507.22139 (2025)}}.

\bibitem{Heung01} T. F. Heung and M. Franz, {\href{https://journals.aps.org/prb/abstract/10.1103/PhysRevB.111.205145}{Phys. Rev. B 111, 205145 (2025)}}.


\bibitem{JKrempasky01} J. Krempask\'y et al., {\href{https://www.nature.com/articles/s41586-023-06907-7}{Nature {\bf 626}, 517-522 (2024)}}.
\bibitem{Lee01} S. Lee et al., {\href{https://journals.aps.org/prl/abstract/10.1103/PhysRevLett.132.036702}{Phys. Rev. Lett. {\bf 132}, 036702 (2024)}}.
\bibitem{Osumi01} T. Osumi et al., {\href{https://journals.aps.org/prb/abstract/10.1103/PhysRevB.109.115102}{Phys. Rev. B {\bf 109}, 115102 (2024)}}.


\bibitem{Ding01} J. Ding et al.,{\href{https://journals.aps.org/prl/abstract/10.1103/PhysRevLett.133.206401}{Phys. Rev. Lett. {\bf 133}, 206401 (2024)}}.
\bibitem{Lu01} W. Lu et al., {\href{https://pubs.acs.org/doi/10.1021/acs.nanolett.5c00482}{Nano Lett. {\bf 25}, 7343 (2025)}}.
\bibitem{CLi01} C. Li et al., {\href{https://www.nature.com/articles/s42005-025-02232-9}{Commun. Phys. {\bf 8}, 311 (2025)}}.



\bibitem{BJiang01} B. Jiang et al., {\href{https://www.nature.com/articles/s41567-025-02822-y}{Nat. Phys. {\bf 21}, 754 (2025)}}.
\bibitem{FZhang01} F. Zhang et al., {\href{https://arxiv.org/abs/2407.19555}{arXiv:2407.19555}}.







\bibitem{Miyasaka01} S. Miyasaka, Y. Okimoto, M. Iwama, and Y. Tokura, {\href{https://journals.aps.org/prb/abstract/10.1103/PhysRevB.68.100406}{Phys. Rev. B {\bf 68}, 100406(R) (2003)}}.


\bibitem{XJZhang01} X.-J. Zhang, E.Koch, and E. Pavarini, {\href{https://journals.aps.org/prb/abstract/10.1103/PhysRevB.106.115110}{Phys. Rev. B {\bf 106}, 115110 (2022)}}.

\bibitem{Wohlfeld01} K. Wohlfeld, A. M. Oles, and P. Horsch, {\href{https://journals.aps.org/prb/abstract/10.1103/PhysRevB.79.224433}{Phys. Rev. B {\bf 79}, 224433 (2009)}}.

\bibitem{Cuono01} G. Cuono, R. M. Sattigeri, J. Skolimowski, and C. Autieri, {\href{https://arxiv.org/abs/2306.17497}{arXiv:2306.17497 (2023)}}.

\bibitem{Smejkal02} L. {\v{S}}mejkal et al.,  {\href{https://journals.aps.org/prl/abstract/10.1103/PhysRevLett.131.256703}{Phys. Rev. Lett. {\bf 131}, 256703 (2023)}}.

\bibitem{Maier01} T. A. Maier and S. Okamoto,  {\href{https://journals.aps.org/prb/abstract/10.1103/PhysRevB.108.L100402}{Phys. Rev. B {\bf 108}, L100402  (2023)}}.
\bibitem{Gaitan01} F. G.-Gaitan, A. Kefayati, J. Q. Xiao, and B. K. Nikoli\'c,  {\href{https://journals.aps.org/prb/abstract/10.1103/PhysRevB.111.L020407}{Phys. Rev. B {\bf 111}, L020407  (2025)}}.
\bibitem{Kaushal02} N. Kaushal and M. Franz,  {\href{https://journals.aps.org/prl/abstract/10.1103/s31h-hk2v}{Phys. Rev. Lett. {\bf 135}, 156502 (2025)}}.


\bibitem{Liu01} Z. Liu, M. Ozeki1, S. Asai, S. Itoh, and T. Masuda, {\href{https://journals.aps.org/prl/abstract/10.1103/PhysRevLett.133.156702}{Phys. Rev. Lett. {\bf 133}, 156702 (2024)}}.
\bibitem{Singh01} A. K. Singh et al., {\href{https://arxiv.org/abs/2511.16086}{arXiv:2511.16086 (2025)}}.

\bibitem{Saitoh01} E. Saitoh, S. Okamoto, K. T. Takahashi, K. Tobe, K. Yamamoto, T. Kimura, S. Ishihara, S. Maekawa, and Y. Tokura, {\href{https://www.nature.com/articles/35065547}{Nature {\bf 410}, 180-183 (2001)}}.

\bibitem{Gruninger01} M. Gr\"uninger, R. R\"uckamp, M. Windt, P. Reutler, C. Zobel, T. Lorenz, A. Freimuth, and A. Revcolevschi, {\href{https://www.nature.com/articles/418039a}{Nature {\bf 418}, 39-40 (2002)}}.


\bibitem{Schlappa01} J. Schlappa, et al. {\href{https://www.nature.com/articles/nature10974}{Nature {\bf 485}, 82-85 (2012)}}.

\bibitem{JLi01} J. Li, et al. {\href{https://journals.aps.org/prl/abstract/10.1103/PhysRevLett.126.106401}{Phys. Rev. Lett. {\bf 126}, 106401 (2021)}}.

\bibitem{Daghofer02} M. Daghofer, K. Wohlfeld, and J. van den Brink, {\href{https://arxiv.org/abs/2506.03261}{arXiv:2506.03261 (2025)}}.

\bibitem{Fishman01} M. Fishman, S. R. White, and E. M. Stoudenmire, {\href{https://scipost.org/SciPostPhysCodeb.4}{SciPost Phys. Codebases, {\bf 4} (2022)}}.


\bibitem{Giuliani01}Giuliani, G. F., and G. Vignale, {\href{https://www.cambridge.org/core/books/quantum-theory-of-the-electron-liquid/EA75F41350A1C41D5E1BD202D539BB9E}{Quantum Theory of the
Electron Liquid, 2005, Cambridge University Press, Cambridge}}.

\bibitem{RColdea01} R. Coldea et al., {\href{https://doi.org/10.1103/PhysRevLett.86.5377} {Phys. Rev. Lett. {\bf 86}, 5377 (2001)}}.



\bibitem{ZXiao01} Z. Xiao, J. Zhao, Y. Li, R. Shindou, and Z.-D. Song, {\href{https://journals.aps.org/prx/abstract/10.1103/PhysRevX.14.031037}{Phys. Rev. X {\bf 14}, 031037  (2024)}}.

\bibitem{Mook01} A. Mook, R. R. Neumann, J. Henk, and I. Mertig, {\href{https://journals.aps.org/prb/abstract/10.1103/PhysRevB.100.100401}{Phys. Rev. B {\bf 100}, 100401(R) (2019)}}.

\bibitem{Rezende01} S. M. Rezende, R. L. R.-Su{\'a}rez, and A. Azevedo, {\href{https://journals.aps.org/prb/abstract/10.1103/PhysRevB.93.014425}{Phys. Rev. B {\bf 93}, 014425 (2016)}}.

\bibitem{QCui01} Qirui Cui, Bowen Zeng, Ping Cui, Tao Yu, and Hongxin Yang, {\href{https://journals.aps.org/prb/abstract/10.1103/PhysRevB.108.L180401}{Phys. Rev. {\bf B} 108, L180401 (2023)}}.

\bibitem{Taniguchi01} Mari Taniguchi, Satoshi Haku, Hyun-Woo Lee, and Kazuya Ando {\href{https://www.nature.com/articles/s41467-025-62703-z}{Nature Communications {\bf 16}, Article number: 8038 (2025)}}.

\bibitem{Gunnink01} Pieter M. Gunnink, Jairo Sinova, Alexander Mook,{\href{https://arxiv.org/abs/2502.18007}{arXiv:2502.18007 (2025)}}.


\bibitem{Lin01} L.-F. Lin, N. Kaushal, Y. Zhang, A. Moreo, and E. Dagotto, {\href{https://journals.aps.org/prmaterials/abstract/10.1103/PhysRevMaterials.5.025001}{Phys. Rev. Materials {\bf 5}, 025001 (2021)}}.

\bibitem{Pandey01} B. Pandey, Y. Zhang, N. Kaushal, R. Soni, L.-F. Lin, W.-J. Hu, G. Alvarez, and E. Dagotto, {\href{https://journals.aps.org/prb/abstract/10.1103/PhysRevB.103.045115}{Phys. Rev. B {\bf 103}, 045115 (2021)}}.

\bibitem{Meier01} Q. N. Meier, A. Carta, C. Ederer, and A. Cano.  {\href{https://arxiv.org/abs/2502.01515}{arXiv:2502.01515 (2025)}}.

\bibitem{Greedan01} J.E. Greedan and W. Gong, {\href{https://www.sciencedirect.com/science/article/abs/pii/092583889290393N}{J. Alloys and Comp. {\bf 180} 281 (1992)}}.

\bibitem{Ricco01} S. Ricc{\`o} et al.,  {\href{https://www.nature.com/articles/s41467-018-06945-0}{Nature Communications {\bf 9}, Article number: 4535 (2018)}}.

\bibitem{SM} Supplementary material,{\href{}{}}.


\end{thebibliography}

\begin{thebibliography}{10}


\bibitem{Colpa01_SM} J.H.P. Colpa, { \href{https://www.sciencedirect.com/science/article/abs/pii/0378437178901607}{Physica A {\bf 93}, 327 (1978)}}.

\bibitem{QCui01_SM} Qirui Cui, Bowen Zeng, Ping Cui, Tao Yu, and Hongxin Yang, {\href{https://journals.aps.org/prb/abstract/10.1103/PhysRevB.108.L180401}{Phys. Rev. {\bf B} 108, L180401 (2023)}}.

\bibitem{Giuliani01_SM}Giuliani, G. F., and G. Vignale, {\href{https://www.cambridge.org/core/books/quantum-theory-of-the-electron-liquid/EA75F41350A1C41D5E1BD202D539BB9E}{Quantum Theory of the
Electron Liquid, 2005, Cambridge University Press, Cambridge}}.

\bibitem{Remund01_SM} K. Remund, R. Pohle, Y. Akagi, J. Romhányi,  and N. Shannon, {\href{https://journals.aps.org/prresearch/abstract/10.1103/PhysRevResearch.4.033106}{Phys. Rev. Research {\bf 4}, 033106 (2022)}}.

\end{thebibliography}
\end{document}